\documentclass[journal,draftcls,onecolumn,12pt,twoside]{IEEEtranTCOM}

\usepackage{graphicx} 
\usepackage{epstopdf} 
\usepackage{cite}

\usepackage{amssymb}
\usepackage[tbtags]{amsmath}
\usepackage[thmmarks,amsmath]{ntheorem} 
\qedsymbol{\ensuremath{\square}} 
{   
	\theoremheaderfont{\bf}
	\theorembodyfont{\rm}
	\theoremseparator{:}
	\newtheorem{XingDef}{Definition}
	\newtheorem{XingRmk}{Remark}
	\newtheorem{XingCorlay}{Corollary}
	\newtheorem{XingPropo}{Proposition}
}

\usepackage{algorithmic}
\usepackage{graphicx}
\usepackage{textcomp}

\usepackage{xcolor}
\usepackage{color}

\usepackage{amssymb}
\usepackage{bbm} 
\usepackage{bm}

\usepackage{algorithm}
\usepackage{algorithmic}
\usepackage{url}
\usepackage{color}

\usepackage{graphicx} 
\usepackage{epstopdf} 

\usepackage{algorithm}
\usepackage{algorithmic}
\usepackage{url}
\usepackage{color}

\usepackage{multirow} 

\usepackage{subfig}

\usepackage{tablefootnote}

\usepackage{caption}

\allowdisplaybreaks[1]

\usepackage{booktabs}
\usepackage{threeparttable}
\usepackage{footnote}
\usepackage{enumerate}

\usepackage{yhmath}

\usepackage{array}

\usepackage{diagbox}

\def\BibTeX{{\rm B\kern-.05em{\sc i\kern-.025em b}\kern-.08em
		T\kern-.1667em\lower.7ex\hbox{E}\kern-.125emX}}
\begin{document}
%
%
%
%

\title{Privacy Leakage in Proactive VR Streaming: Modeling and Tradeoff}
\author{\IEEEauthorblockN{Xing Wei, Chenyang Yang, and Chengjian Sun}
}

%
%


\maketitle

\vspace{-2cm}
\begin{abstract}
	Proactive tile-based virtual reality (VR) video streaming employs the viewpoint of a user to predict the tiles to be requested, renders and delivers the predicted tiles before playback.
	Recently, it has been found that the identity and preference of the user can be inferred from the  trace of viewpoint uploaded for proactive streaming, which indicates that viewpoint leakage incurs privacy leakage. In this paper, we strive to answer the following questions regarding viewpoint leakage during proactive VR video streaming. When is the viewpoint leaked? Can privacy-preserving approaches (e.g., federated or individual training, using predictors with no need for training, or predicting locally) avoid viewpoint leakage? We find that if the prediction error or the quality of experience (QoE) metric is uploaded for adaptive streaming, the real viewpoint can be inferred even with the privacy-preserving approaches. Then, we define \textit{viewpoint leakage probability} to characterize the accuracy of the inferred viewpoint, and respectively derive the probability when uploading prediction error and QoE metric. We find that the viewpoint leakage probability can be reduced by sacrificing QoE or increasing resources.
	Simulation with the state-of-the-art predictor over a real dataset shows that such a tradeoff does not exist only in rare cases.
\end{abstract}

\vspace{-0.35cm}
\begin{IEEEkeywords}
	\vspace{-0.35cm}
	Privacy-preserving VR, viewpoint leakage, proactive VR video streaming
\end{IEEEkeywords}

%

\vspace{-0.35cm}
\section{Introduction}
\vspace{-1mm}
As the main type of wireless virtual reality (VR) services, 360$^\circ$ video streaming consumes a large amount of computing and communication resources to avoid playback stalls and black holes that degrade the quality of experience (QoE) \cite{survey_Hsu,VR_IoT}. However, humans can only see a small spherical cap of the full panoramic sphere (e.g., 18\% of the sphere\cite{NTHU_dataset}) at arbitrary time, which is called field of view (FoV). Considering that the center of the FoV (i.e., viewpoint) is predictable, proactive VR video streaming is proposed to improve the QoE with less resources \cite{optimizing_VR}, which has been investigated intensively.
In \cite{VR_TWC,VR_mobile_computing,yingcui_TWC,VR_broadcast_TWC,VR_transcoding_JSAC,VR_letter,changyangshe_VR_globecom,junnizou_TCSVT},
communication, computing, or caching resource allocation was optimized in various scenarios. In \cite{Xueshihou_openjournal,TRACK}, viewpoint prediction methods were proposed and evaluated.	
In \cite{VR_learning_based_TWC,VR_RIS_JSAC,LSTM_update,QoE_feedback_2,JSAC_private_VR,Xing_VR_Shannon}, both viewpoint prediction and resources allocation were studied. The QoE depends on the performance of  prediction and the configured resources for communication and computing. To keep the QoE stable during video streaming, the instantaneous prediction error or the QoE should be uploaded for adaptive streaming \cite{junnizou_TCSVT,VR_survey_prediction,QoE_feedback_2,changyangshe_VR_globecom}.

\vspace{-4mm}\subsection{Motivation and Major Contributions}
Proactive VR video streaming contains two stages: predictor training and online streaming. Predictor training  can be operated at the multi-access edge computing (MEC) server or the head mounted display (HMD), and is unnecessary for some simple predictors as those evaluated in \cite{optimizing_VR,TRACK}. Online streaming contains viewpoint predicting, proactive rendering and delivering.

Most existing works in literature neglect a fact that privacy may be leaked during  proactive VR video streaming, where the request of a VR video and the trace of viewpoint should be uploaded. Although the request of a VR video can be anonymous to avoid privacy leakage, as the key information to improve QoE and save resources, the trace of viewpoint is inevitable to be uploaded. Then, will privacy be leaked from the viewpoint trace?

Recent works reveal the risk. With less than five minutes' viewpoint trace, the identity of 95\% users among all the 511 users can be correctly identified as demonstrated in \cite{privacy_VR_identifiability}.
Furthermore, with the trace of viewpoint, the content in the FoV that a user chooses to see is also leaked. This can be used to infer the intent and preference of the user \cite{privacy-preserving_eye_tracking_2021}. Therefore, viewpoint leakage indeed incurs privacy leakage.

When considering viewpoint leakage in proactive VR video streaming procedure, several fundamental questions arise:
\begin{itemize}
	\item Can existing privacy-preserving approaches, e.g., federated learning in predictor training stage and local predicting in online streaming stage, avoid viewpoint leakage?
	\item If the viewpoint leakage is unavoidable, when does the most serious leakage happens, and
	when does the most minor leakage happens?
	\item Which factors affect viewpoint leakage? Is there any relation between viewpoint leakage, QoE, and configured resources? 
\end{itemize}

In this paper, we strive to answer these questions. Our main contributions are as follows.
\begin{itemize}
	\item We find that although the viewpoint leakage can be avoided with privacy-preserving approaches in predictor training, the real viewpoint can be inferred during online streaming.
	\item To characterize the leakage, we define \textit{viewpoint leakage probability} and derive the probability when uploading the prediction error and the QoE metric, from which we find the conditions that the probability achieves the maximum or minimum.
	\item We find that viewpoint leakage probability depends on the prediction error and configured resources. Moreover, there is a privacy-QoE tradeoff and a privacy-resources tradeoff, where to reduce the leakage probability, either the QoE has to be scarified or more communication and computing resources should be configured. Simulations with the state-of-the-art predictor over a real dataset show that the tradeoffs exist only except rarely happened cases when uploading the QoE metric.
\end{itemize}

\vspace{-4mm}\subsection{Related Works}
As far as the authors known, there are no prior works to investigate viewpoint leakage in proactive VR video streaming.

Existing works in the field of computer science propose to protect real gaze position (another useful data for viewpoint prediction)  by adding noise in spatial domain \cite{privacy-preserving_eye_tracking_2021,privacy_def_eye_track,Privacy_Preserving_Gaze_Estimation,Differential_Privacy} or reducing the samples of uploaded gaze position in temporal domain \cite{privacy-preserving_eye_tracking_2021,PrivacEye}. 
However, gaze position may not always be useful for viewpoint prediction. In fact, the state-of-the-art accuracy of viewpoint prediction on real dataset \cite{NTHU_dataset} can be achieved only with the trace of viewpoint \cite{TRACK}. Moreover, the real gaze position (and also the real viewpoint) can be stored at the HMD and hence is unnecessary to protect when the privacy-preserving approaches (say federated learning) are used. \textit{Existing works never consider the fact that the gaze position or viewpoint can still be inferred with the privacy-preserving approaches}. Besides, Existing works consider the gaze position on two dimensional (2D) plane. Yet the actual gaze position or viewpoint is on the sphere, the impact of which on the privacy leakage has never been investigated.

There is only one related work in the field of wireless communications, which considers federated training and local predicting for privacy-preserving \cite{JSAC_private_VR} but does not study if the viewpoint can really be protected.

\vspace{-2mm}
\subsection{Outline}
\vspace{-2mm}

The rest of this paper is organized as follows. Section II introduces the system model, the procedure of proactive VR video streaming, and the definitions of the prediction error and QoE metric. Section III analyzes the viewpoint leakage issue and defines viewpoint leakage probability. Section IV and V derive and analyze the leakage probability when uploading the prediction error and the QoE metric, respectively.
Trace-driven simulation results are provided in Section VI.
Section VII concludes the paper.

\vspace{-2mm}\section{System Model}\label{section:system_model}
Consider a proactive VR video streaming system serving $K$ users with a MEC server, which accesses a VR video library by local caching or high-speed backhaul, thus the delay from the Internet to the server can be ignored. The server is also equipped with powerful computing units for training, predicting, and rendering.

Each user wears an HMD, which can measure and upload the trace of viewpoint,\footnote{Since tile requests can be transformed into the viewpoints \cite{Fixation_Prediction}, we do not consider the case that uploading the tile requests. } calculate and upload the value of real-time QoE or prediction error, and pre-buffer segments. The HMD can also equip with a light-weighted computing unit for training a predictor and predicting the viewpoint. During the playback, each user can turn around freely to view FoVs.

Each $360^{\circ}$ video consists of $L$ segments, each segment consists of $N_f$ panoramic video frames in temporal domain, and each frame consists of $M$ tiles in spatial domain.
For ease of analysis, we assume that the areas of all the tiles within a FoV and a streamed field of view (SFoV) in a panoramic frame can be respectively approximated as the area of the FoV and SFoV \cite{yingcui_TWC}. This can be achieved by sufficiently fine grained tiling or adaptive tiling \cite{junnizou_TCSVT}.

\begin{figure}[htbp]
	\vspace{-0.5cm}
	\centering
	\begin{minipage}[t]{0.77\linewidth}
		\includegraphics[width=0.9\textwidth]{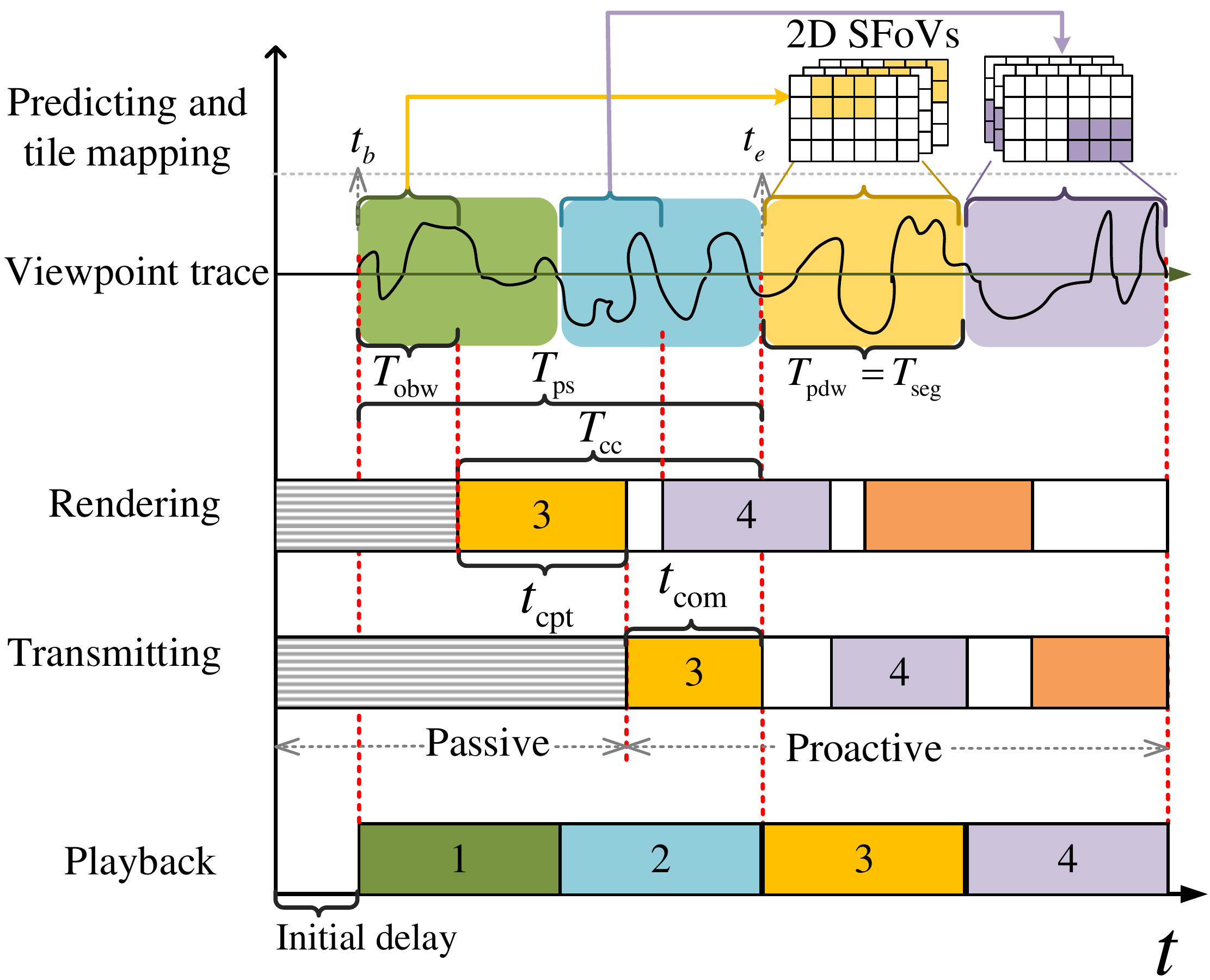}
	\end{minipage}
	\vspace{-0.3cm}
	\caption{Streaming the first four segments of a VR video. $t_b$ is the start time of the observation window, $t_e$ is the start time of playback of the $l_0$th segment, $l_0=3$. }
	\label{Fig:proactive_streaming}
	\vspace{-0.7cm}
\end{figure}

\vspace{-0.3cm}
\subsection{Proactive VR Video Streaming Procedure}\label{subsection:online_streaming_procedure}

To predict the viewpoints, if a predictor needs to be trained, the proactive streaming procedure contains two stages: (1) offline predictor training, and (2) online video streaming. Otherwise, stage (1) can be omitted (say for the simple predictors used in \cite{optimizing_VR,TRACK}).

Predictor training can be operated at the MEC server or the HMD. When training at the server (i.e., centralized learning), the real viewpoint traces of each user should be uploaded. When training at each HMD (i.e., federated learning), the real viewpoint traces can be stored locally and only the model parameters of the predictor are uploaded to the server. When training local predictors individually at each HMD, both the real viewpoint traces and the model parameters of predictors can be stored at the HMD.

The procedure in the online streaming stage is shown in Fig. \ref{Fig:proactive_streaming}. When a user requests a VR video, the MEC server first renders and transmits the initial ($l_{0}-1$) segments in a passive manner \cite{transmission_mode-standard-update}. After an initial delay, the first segment begins to play at the time instant $t_b$, i.e., the start time of an observation window. Then, proactive streaming begins. In the sequel, we take the $l_0$th segment as an example for elaboration.

After the viewpoint trace in the observation window with duration $T_{\mathrm{obw}}$ is collected (i.e., at the end time of the window), the viewpoint sequence in a prediction window with duration $T_{\mathrm{pdw}}=T_{\mathrm{seg}}$ can be predicted at the MEC server or a HMD, and the computing and communication resources can be configured at the MEC. According to the distance between the predicted viewpoint and the center of each tile,
the MEC server can determine the tiles for the $N_f$ frames of the segment to be streamed \cite{Fixation_Prediction,Xing_VR_Shannon}. Then, the server renders these tiles with duration $t_{\mathrm{cpt}}$ to generate the $N_f$ images of SFoVs, which are then transmitted with duration $t_{\mathrm{com}}$. To avoid stalling, $N_f$ SFoVs in the $l_0$th segment should be delivered before the playback start time of the $l_0$th segment, i.e., the time instant $t_e$. The duration from $t_b$ to $t_e$ is the online proactive streaming time for a segment, denoted as $T_{\mathrm{ps}}$. We can observe from the figure that $T_{\mathrm{ps}}=(l_0 -1)T_{\mathrm{seg}}$, where in the example $l_{0}=3$ and hence $T_{\mathrm{ps}}=2T_{\mathrm{seg}}$. The durations for observation, computing, and transmission should satisfy $T_{\mathrm{obw}} + T_{\mathrm{cc}}=T_{\mathrm{ps}}$, where $T_{\mathrm{cc}}\triangleq t_{\mathrm{com}} + t_{\mathrm{cpt}}$ is the total duration for communication and computing. A predictor can be more accurate with a smaller value of $T_{\mathrm{cc}}$. This is because the viewpoints to be predicted are closer to and hence are more correlated with the viewpoint sequence in the observation window \cite{Xing_VR_Shannon}. Given a predictor and the required viewpoint prediction accuracy,
the value of $T_{\mathrm{cc}}$ can be pre-determined \cite{Xing_VR_Shannon,Fixation_Prediction}.

During the playback of each segment, the instantaneous QoE metric or the instantaneous prediction error can be calculated at the HMD and uploaded to the MEC server for adaptive streaming, i.e., when the prediction error or the QoE is unacceptable, the server streams more tiles \cite{junnizou_TCSVT,VR_survey_prediction,QoE_feedback_2,changyangshe_VR_globecom}.

\vspace{-2mm}\subsection{Field of View on Sphere}

\begin{figure}[htbp]
	\vspace{-0.75cm}
	\centering
	\subfloat[FoV]{\label{Fig:Viewport_on_Sphere}
		\begin{minipage}[c]{0.33\linewidth}
			\centering
			\includegraphics[width=1\textwidth]{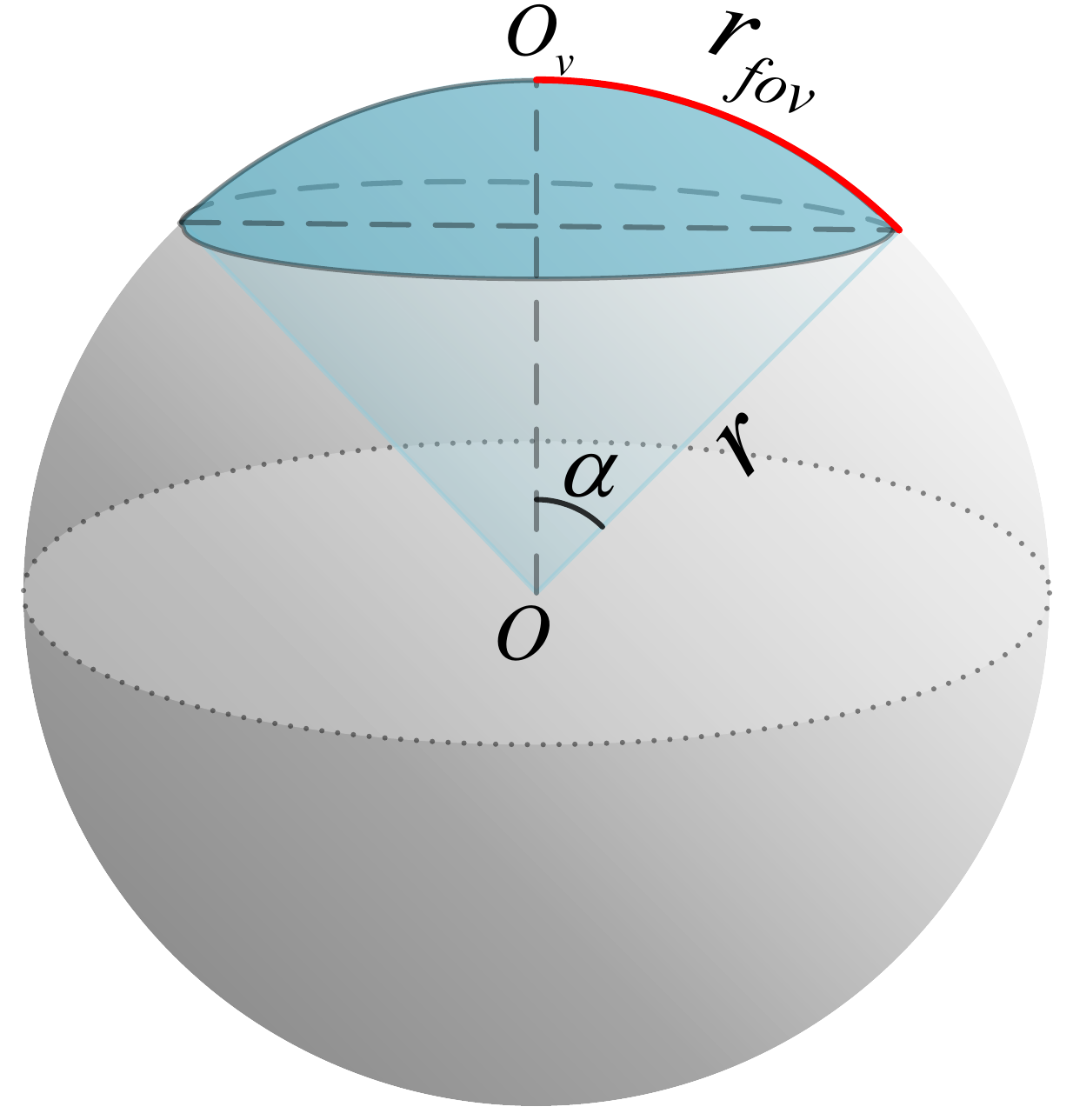}
		\end{minipage}
	}
	\subfloat[SFoV]{\label{Fig:pred_Viewport_on_Sphere}
		\begin{minipage}[c]{0.33\linewidth}
			\centering
			\includegraphics[width=1\textwidth]{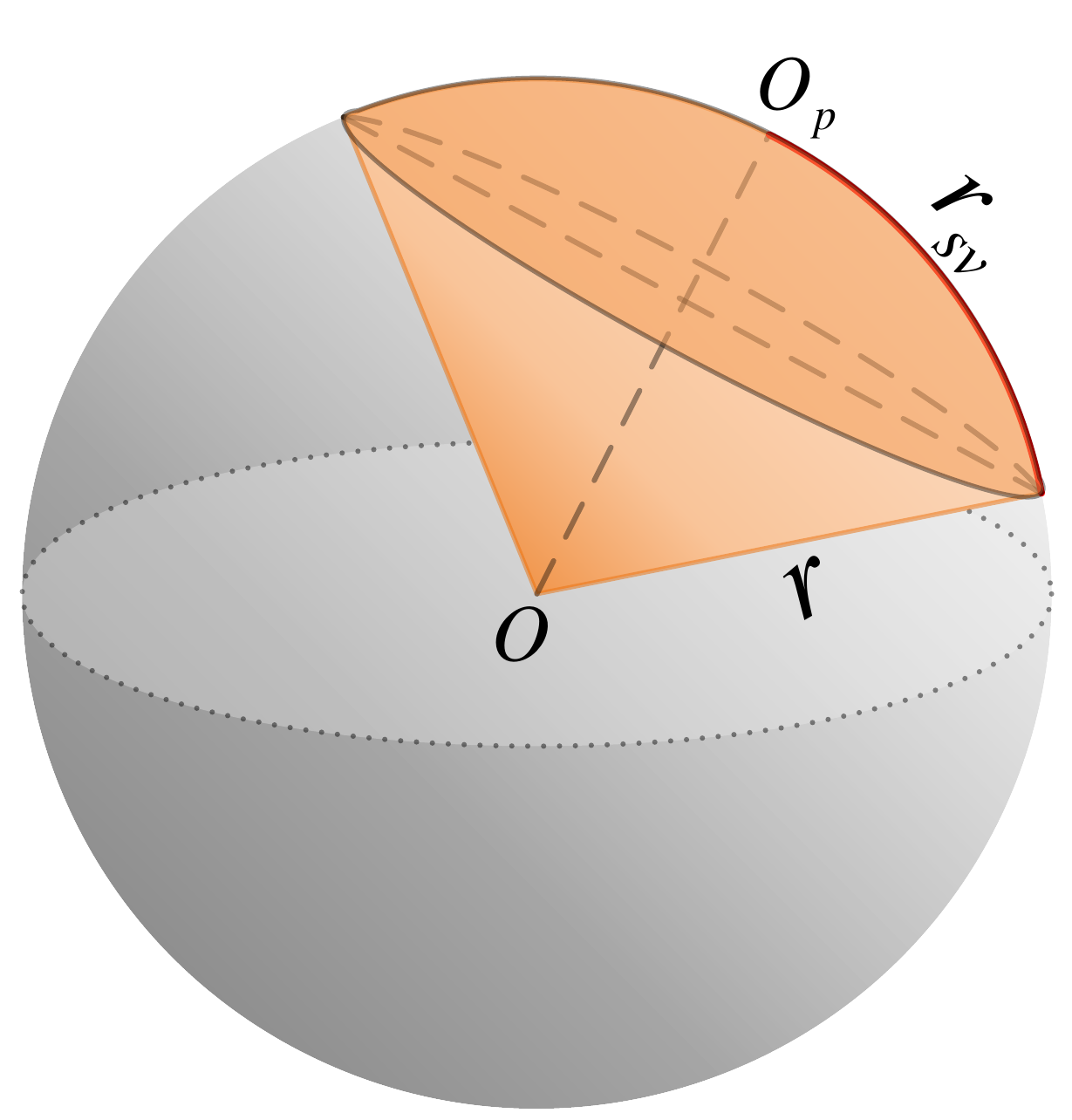}
		\end{minipage}
	}
	\subfloat[$A_{\textit{ol}}$ (red region)]{\label{Fig:qoe_predict_error}
		\begin{minipage}[c]{0.33\linewidth}
			\centering
			\includegraphics[width=1\textwidth]{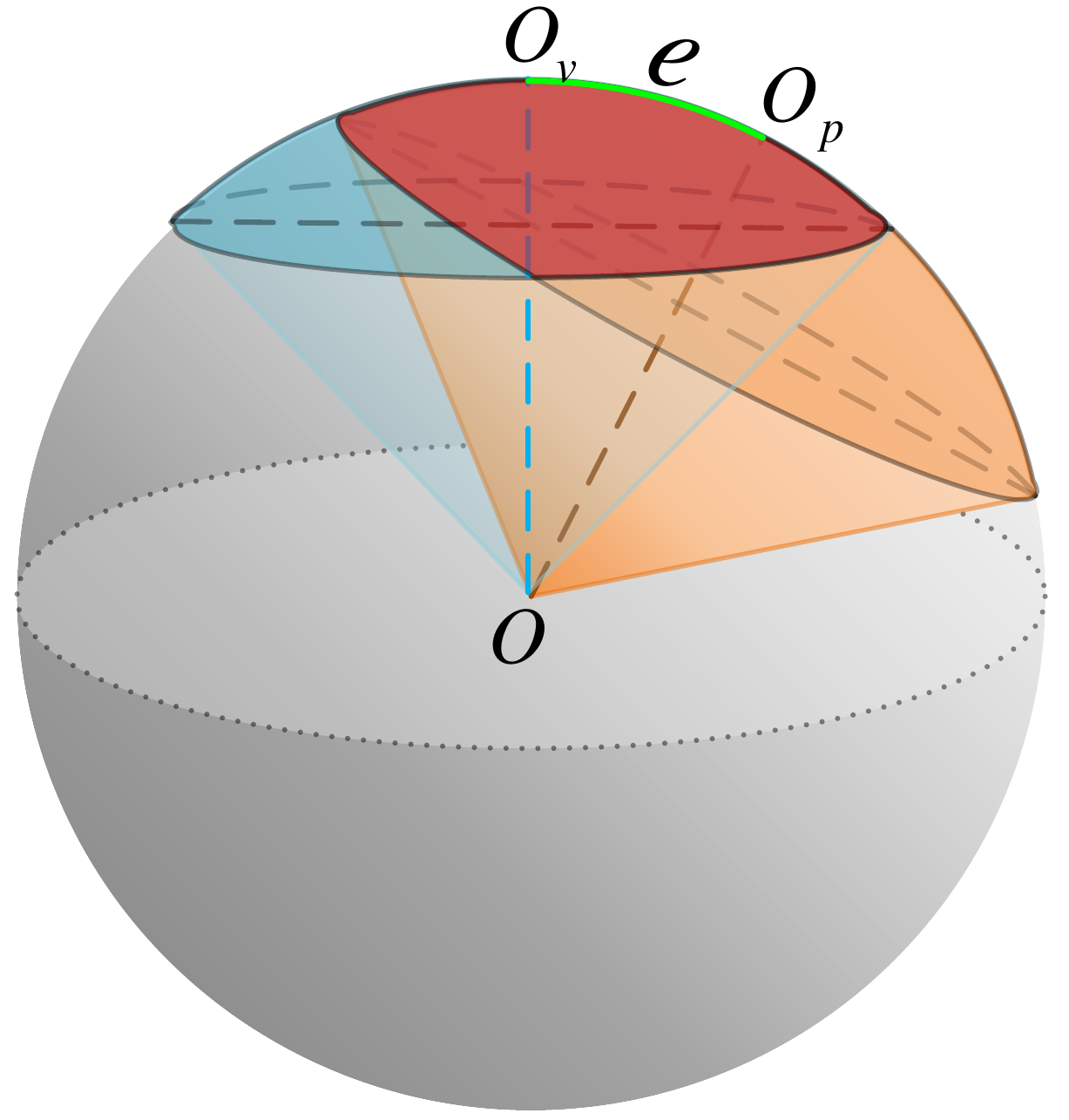}
		\end{minipage}
	}
	\vspace{-0.3cm}
	\caption{FoV, SFoV, and QoE on the unit sphere, $O_v$ and $O_p$ are respectively the real and predicted viewpoint, $r_{\textit{fov}}$ and $r_{\textit{sv}}$ are respectively the radius of FoV and SFoV, $e$ is the prediction error.}\label{Fig:FoV_SFoV_sphere}	
	\vspace{-0.65cm}
\end{figure}

When watching a panoramic video, the user wearing an HMD is at the center of the unit sphere, i.e., $O$, as illustrated in Fig. \ref{Fig:Viewport_on_Sphere}. The FoV of the HMD can be considered as a \textit{spherical cap} of the sphere \cite{NTHU_dataset,Fixation_Prediction
,TRACK}. For any given HMD, the size of FoV is determined.
Denote viewpoint as $O_v$. The spherical distance from $O_v$ to the base of the cap is $r_{\textit{fov}}$ (in \textit{radian}), and is referred to as the ``cap radius". The half apex angle of the cone corresponding the cap is $\alpha$. When measures in radian, $\alpha=\frac{r_{\textit{fov}}}{r}=r_{\textit{fov}}$. Then, the area of the FoV is $	A_{\textit{fov}} = 2\pi \left(1 - \cos (\alpha)\right) = 2\pi \left(1 - \cos (r_{\textit{fov}})\right)$.

\vspace{-2mm}\subsection{Computing and Transmission Model}

\subsubsection{Computing Model}
For each user, the number of bits that can be rendered per second, referred to as the computing rate \cite{Xing_VR_Shannon}, is $c_{\mathrm{cpt},k} \triangleq \frac{\mathcal{F}_{\mathrm{cpt}}}{K\cdot\mu_r}$ (in bit/s),
where $\mathcal{F}_{\mathrm{cpt}}$ is the configured resource at the server for rendering  (in floating-point operations per second, FLOPs/s),
$\mu_r$ is the required floating-point operations (FLOPs) for rendering one bit of FoV (in FLOPs/bit) \cite{Xing_VR_Shannon}.

\subsubsection{Transmission Model}
The base station co-located with the server equipped with $N_t$ antennas serves $K$ single-antenna users using zero-forcing beamforming.
The instantaneous data rate in $t_{\mathrm{com}}$ for the $k$th user is $	c_{\mathrm{com},k}=B\log_2 \left(1+\frac{p_k d_k^{-\beta}|(\mathbf{h}_k)^H\mathbf{w}_k|^2}{\sigma^2} \right)$,
where $B$ is the bandwidth, $\mathbf{h}_k$ and $\mathbf{w}_k$ are respectively the channel vector and beamforming vector, $d_k$ is the distance from the BS to the user, $p_k$ is the transmit power, $\beta$ is the path-loss exponent, $\sigma^2$ is the noise power, and $(\cdot)^{H}$ denotes conjugate transpose. With duration $t_{\mathrm{com}}$, the number of bits in a segment that can be protectively transmitted is $\overline{c}_{\mathrm{com},k}t_{\mathrm{com}}$, where $\overline{c}_{\mathrm{com},k}$ is the time-average data rate over $t_{\mathrm{com}}$. Since the future channels of each user are unknown when the communication resource is allocated at the end of the observation window, we use ensemble-average rate $\mathbb{E}_h\{c_{\mathrm{com},k}\}$ to approximate $\overline{c}_{\mathrm{com},k}$ \cite{Xing_VR_Shannon}, where $\mathbb{E}_h\{\cdot\}$ is the expectation over $h$.

Since the resources are configured orthogonally among $K$ users, we consider an arbitrary user and omit the subscript $k$ in the computing rate and ensemble-average data rate in the sequel.

\subsubsection{Capability for Streaming}
We use the ratio of tiles in a segment that can be rendered and transmitted with configured resources (reflected implicitly by the data rate and computing rate) to measure the capability of the system for streaming tiles, which is \cite{Xing_VR_Shannon}
\begin{align}\label{C_cc_max}
	C_{} = \min\left\{\frac{T_{\mathrm{cc}}}{N_f M (\frac{s_{\mathrm{com}}}{\mathbb{E}_h\{c_{\mathrm{com}}\}} + \frac{s_{\mathrm{cpt}}}{c_{\mathrm{cpt}}})} , 1 \right\}\in[0,1]
\end{align}
where $s_{\mathrm{com}} = {px}_w \cdot{px}_h \cdot b /\gamma_{c}$ \cite{HuaWei_Cloud_VR} is the number of bits in a tile for transmission, $s_{\mathrm{cpt}} = {px}_w \cdot{px}_h \cdot b $ is the number of bits in a tile for rendering, ${px}_w$ and ${px}_h$ are the pixels in width and height of a tile, $b$ is the number of bits per pixel relevant to color depth, and $\gamma_c$ is the compression ratio of video.

The number of tiles within SFoV is identical among $N_f$ frames in a segment, hence the ratio of the SFoV to the panoramic frame is also $C$, i.e., $C = \frac{A_{\textit{sv}}}{4\pi}$,
where $A_{\textit{sv}}$ denotes the area of the SFoV, and $4\pi$ is the surface area of the unit sphere.

\subsection{Streamed Field of View on Sphere} 
Taking the predicted viewpoint $O_p$ as the center, SFoV contains all the tiles whose spherical distances from their centers to $O_p$ are no more than a given value, thus we can assume that the SFoV is also a spherical cap with center $O_p$, as shown in Fig. \ref{Fig:pred_Viewport_on_Sphere}. Then, the area of the SFoV is
\begin{align}\label{S_pred_FoV}
	A_{\textit{sv}} = 2\pi \Big(1 - \cos \big(r_{\textit{sv}}\big)\Big)
\end{align}
where $r_{\textit{sv}}$ is the cap radius of the SFoV.

By substituting \eqref{C_cc_max} and $C = \frac{A_{\textit{sv}}}{4\pi}$ into \eqref{S_pred_FoV}, we obtain the radius of SFoV as
\begin{align}\label{r_str_com_relation}
r_{\textit{sv}} = \arccos \left( 1 - 2C \right)= \arccos \left( 1 - 2\min\left\{\frac{T_{\mathrm{cc}}}{N_f M (\frac{s_{\mathrm{com}}}{\mathbb{E}_h\{c_{\mathrm{com}}\}} + \frac{s_{\mathrm{cpt}}}{c_{\mathrm{cpt}}})} , 1 \right\} \right)\in[0,\pi]
\end{align}
\begin{XingRmk}\label{Remark:r_sv_is_resources}
As shown in \eqref{r_str_com_relation}, $r_{\textit{sv}}$ respectively increases with $\mathbb{E}_h\{c_{\mathrm{com}}\}$ or $c_{\mathrm{cpt}}$. Hence, $r_{\textit{sv}}$ can be used to reflect the amount of resources configured for VR video streaming.
\end{XingRmk}
\vspace{-0.5cm}

When $r_{\textit{sv}}=\pi$, according to \eqref{r_str_com_relation} and \eqref{S_pred_FoV},
$C=1$ and $A_{\textit{sv}}=4\pi$, which indicates that the complete panoramic sphere is streamed. In this case, proactive VR streaming degrades into passive streaming and prediction is unnecessary \cite{transmission_mode-standard-update}. When $r_{\textit{sv}}=0$, $C=0$ and $A_{\textit{sv}}=0$, which indicates that no resources are configured and the panoramic frame is not streamed at all.

\subsection{Spherical Distance as the Prediction Error}
Spherical distance, known as orthodromic distance, has been considered as the most appropriate metric to measure the viewpoint prediction error of FoV on the unit sphere \cite{TRACK}. 
As shown in Fig. \ref{Fig:qoe_predict_error}, the spherical distance from the real viewpoint $O_v=(\theta_v, \varphi_v)$ to the predicted viewpoint $O_p=(\theta_p, \varphi_p)$ can be expressed as
\begin{align}\label{spherical_dist_def}
	e \triangleq \mathtt{d}(O_v, O_p) =  \arccos \left(\cos(\varphi_v)\cos(\varphi_p)\cos(|\theta_v - \theta_p|) + \sin(\varphi_v)\sin(\varphi_p) \right) \in[0,\pi]
\end{align}
where $\mathtt{d}(x, y)$ denotes the spherical distance between two points $x$ and $y$,
$\theta$ and $\varphi$ are respectively the longitude and latitude of a point on a unit sphere \cite{TRACK,junnizou_TSTSP}.
When $e=0$, the predicted and real viewpoints coincide. When $e=\pi$, the prediction error reaches the maximum.

After the user watches a video frame of a segment, the instantaneous prediction error can be calculated from \eqref{spherical_dist_def} at the HMD.

\subsection{Quality of Experience}
For proactive VR video streaming, there will be no latency if the SFoVs in a segment can be delivered before the playback of the segment. However, due to the prediction error or insufficient resources, black holes may appear \cite{survey_Hsu}. We define the correctly streamed portion in a FoV as the QoE metric \cite{Xing_VR_Shannon}, i.e.,  the ratio of the overlapped area of FoV and SFoV to the area of FoV, as shown in Fig. \ref{Fig:qoe_predict_error}. Denote the overlapped area of FoV and SFoV as $A_{\textit{ol}}\in\left[0, \min\{A_{\textit{fov}}, A_{\textit{sv}}\}\right]$, then the QoE metric can be expressed as
\begin{align}\label{QoE_def}
	\mathrm{QoE}\triangleq \frac{A_{\textit{ol}} }{A_{\textit{fov}}}\in[0,100\%]
\end{align}
which is larger when the prediction error is smaller \cite{Xing_VR_Shannon}. After the user watches a video frame of a segment, the value of $\mathrm{QoE}$ can be calculated at the HMD.

\section{Viewpoint Leakage Probability in Proactive VR Video Streaming}

In this section, we show that although real viewpoint is unnecessary to be uploaded during proactive VR video streaming, it can still be inferred from the uploaded viewpoint prediction and the prediction error or the QoE metric. Then, we define the viewpoint leakage probability.

\vspace{-2mm}\subsection{Viewpoint Leakage in Proactive Streaming Procedure}\label{subsection:Viewpoint_Leakage_Procedure}

During the proactive streaming procedure, the viewpoint may be leaked out in the predictor training or the online streaming stage. According to whether a predictor requires training and where the training and predicting are respectively executed, we provide seven cases in Table \ref{table:FoV_leakage_one_manner}.

\begin{table}[htbp]
	\vspace{-0.7cm}
	\caption{Uploaded data from HMD in proactive VR video streaming procedure}\label{table:FoV_leakage_one_manner}
	\vspace{-0.6cm}
	\begin{center}
		\begin{tabular}{|c|c|c|c|c|c|c|}
			\hline
			\multirow{2}{*}{No.}&Where&Where&\multirow{2}{*}{Predictor training}&\multicolumn{3}{c|}{Online streaming} \\
			\cline{5-7}
			&to train&to predict&&Viewpoint observation&Viewpoint prediction&Adaptive streaming\\
			\cline{1-7}
			1&MEC&MEC&Real viewpoint& Real viewpoint& \backslashbox[1.7cm]{}{} &\\
			\cline{1-6}
			2&MEC&HMD&Real viewpoint& \backslashbox[1.7cm]{}{}&Predicted viewpoint& \\
			\cline{1-6}
			3&\texttt{HMD}$^{*}$ &HMD& \backslashbox[1.7cm]{}{}&\backslashbox[1.7cm]{}{}  & Predicted viewpoint&Prediction error  or \\
			\cline{1-6}
			\textbf{4}&HMD&HMD&Model parameters &\backslashbox[1.7cm]{}{}& Predicted viewpoint&QoE metric\\
			\cline{1-6}
			5 &HMD&MEC&Model parameters & Real viewpoint&\backslashbox[1.7cm]{}{}& \\
			\cline{1-6}
			6&No need&MEC& \backslashbox[1.7cm]{}{} & Real viewpoint&\backslashbox[1.7cm]{}{} & \\
			\cline{1-6}
			7&No need&HMD& \backslashbox[1.7cm]{}{} & \backslashbox[1.7cm]{}{}&Predicted viewpoint&\\
			\hline
		\end{tabular}
	\end{center}
	\vspace{-0.2cm}
	\footnotesize{ \ \ \ $^*$ This is training individually at each HMD for local predictors. Other cases that training at the HMD are federated.}
	\vspace{-0.6cm}
\end{table}

We can find that privacy-preserving approaches (i.e., training in federated manner \cite{JSAC_private_VR}, training individually at each HMD, selecting training-free predictor\cite{optimizing_VR,TRACK}, or predicting locally \cite{JSAC_private_VR}) can avoid the real viewpoint leakage in the predictor training stage and viewpoint observation of online streaming stage. Nevertheless, the predicted viewpoint needs to be uploaded for viewpoint prediction in the online streaming stage. With the uploaded instantaneous prediction error or QoE metric, the real viewpoint may still be inferred. For example, when $e=0$ or $\mathrm{QoE}=100\%$ and the area of SFoV equals to the area of FoV (i.e., $r_{\textit{sv}}=r_{\textit{fov}}$), one can infer that the predicted viewpoint is the real viewpoint, because the prediction error or QoE indicates the accuracy of the predicted viewpoint.
Then, a natural question is, with the predicted viewpoint at hand, with what probability the real viewpoint can be inferred from the prediction error or the QoE?

\vspace{-3mm}\subsection{$\varepsilon$-Viewpoint Leakage Probability}

We first define the viewpoint-sensitive neighborhood, viewpoint leakage event, and the possible viewpoint zone. Denote $\mathcal{V}$ as the set of all points on a unit sphere.

\vspace{-0.5cm}
\begin{XingDef}
$\varepsilon$-viewpoint-sensitive neighborhood $\mathcal{N}(O_v, \varepsilon)$: A subset of $\mathcal{V}$ where the spherical distance between every point in the subset and the real viewpoint $O_v$ is no larger than $\varepsilon$, i.e., $\mathcal{N}(O_v, \varepsilon)\triangleq \{x|\mathtt{d}(x, O_v)\leq \varepsilon \ \mathrm{and}\  x  \in \mathcal{V}  \}$,
where
$\varepsilon\in[0,r_{\textit{fov}}]$. 
\end{XingDef}
\vspace{-0.5cm}

$\mathcal{N}(O_v, \varepsilon)$ is a region within FoV that a user is not willing to be leaked. The value of $\varepsilon$ reflects the privacy requirement of the user. When $\varepsilon=0$, the user has no privacy requirement. When $\varepsilon=r_{\textit{fov}}$, the whole FoV is required not to be leaked.
We refer to $\varepsilon$ as ``the radius of the viewpoint-sensitive neighborhood".
\vspace{-0.5cm}
\begin{XingDef}
Possible viewpoint zone $\mathcal{Z}_v$: A subset of $\mathcal{V}$ that consists of all possible inferred viewpoints, given the predicted viewpoint $O_p$ and the prediction error or the QoE.
\end{XingDef}
\vspace{-0.5cm}

If $e=0$ or $\mathrm{QoE}=100\%$ and $r_{\textit{sv}}=r_{\textit{fov}}$, the only one inferred viewpoint is the real viewpoint and $\mathcal{Z}_v=\{O_v\}$. If the prediction error or the QoE cannot provide any useful information, the possibly inferred viewpoint can be arbitrary one viewpoint on the unit sphere and $\mathcal{Z}_v=\mathcal{V}$.
\vspace{-0.5cm}
\begin{XingDef}
	$\varepsilon$-viewpoint leakage event: When inferring the real viewpoint, the inferred viewpoint falls in the $\varepsilon$-viewpoint-sensitive neighborhood, i.e., $\widehat{O}_v\in \mathcal{N}(O_v, \varepsilon)$, where $\widehat{O}_v$ is the inferred viewpoint.
\end{XingDef}
\vspace{-0.5cm}

\vspace{-0.5cm}
\begin{XingDef}\label{def_VLP_inital}
\it{$\varepsilon$-viewpoint leakage probability: The probability that $\varepsilon$-viewpoint leakage event happens, which is
\begin{align}\label{def:PL_global}
	\mathrm{Pr}\left\{\widehat{O}_v\in \mathcal{N}(O_v, \varepsilon)\right\} = \min\left\{\frac{\lambda \left[\mathcal{N}(O_v, \varepsilon)\right]}{\lambda [\mathcal{Z}_v]},1\right\}
\end{align}
where $\lambda[\mathcal{X}]$ is the Lebesgue measure of $\mathcal{X}$. When $\mathcal{X}$ is a set of all viewpoints in a curve or a surface, $\lambda[\mathcal{X}]$ is the length of the curve or the area of the surface.
}
\end{XingDef}
\vspace{-0.5cm}

When the prediction error or the QoE can provide sufficient information such that the possible viewpoint zone is limited in the $\varepsilon$-viewpoint-sensitive neighborhood (i.e., $\lambda [\mathcal{Z}_v]\leq \lambda \left[\mathcal{N}(O_v, \varepsilon)\right]$), the real viewpoint is $\varepsilon$-leaked (i.e., $\mathrm{Pr}\left\{\widehat{O}_v\in \mathcal{N}(O_v, \varepsilon)\right\}=1$).

When the prediction error or the QoE cannot provide any useful information such that the possible viewpoint zone is the whole sphere, the real viewpoint can be fully protected. In this case, the measure of the possible viewpoint zone is the surface area of the sphere (i.e., $\lambda[\mathcal{Z}_v]=4\pi$), the measure of viewpoint sensitive neighborhood is the area of a spherical cap with center $O_v$ and radius $\varepsilon$ (i.e., $\lambda[\mathcal{N}(O_v, \varepsilon)] = 2\pi(1 -\cos(\varepsilon) )$), and the viewpoint leakage probability achieves its global minimum, which is $\mathrm{Pr}^{\min} = \frac{1-\cos(\varepsilon)}{2}$.
\vspace{-0.3cm}
\begin{XingRmk}\label{remark2}
$\mathrm{Pr}\left\{\widehat{O}_v\in \mathcal{N}(O_v, \varepsilon)\right\}$ is a monotonically increasing function of $\varepsilon$.
\end{XingRmk}
\vspace{-0.5cm}

This is because $\mathrm{Pr}\big\{\widehat{O}_v\in \mathcal{N}(O_v, \varepsilon)\big\}$ is a monotonically increasing function of $\lambda[\mathcal{N}(O_v, \varepsilon)]$, and $\lambda[\mathcal{N}(O_v, \varepsilon)]$ monotonically increases with $\varepsilon$. Remark \ref{remark2} indicates that the viewpoint privacy will be harder to be protected if the required viewpoint-sensitive neighborhood is larger.

In the sequel, we derive and analyze the $\varepsilon$-viewpoint leakage probability when uploading prediction error and QoE metric, respectively.

\section{Viewpoint Leakage Probability when Uploading Prediction Error}\label{section:Upload_Predicted_Viewpoint_and_Prediction_Error}
In this section, we derive the leakage probability when uploading prediction error. We find a conditional
tradeoff between reducing the leakage probability and improving QoE, while the targets of satisfying the viewpoint privacy requirement and maximizing QoE are always contradictory.

\vspace{-2mm}\subsection{$\varepsilon$-Viewpoint Leakage Probability}\label{subsection:VLP_given_e}
\begin{figure}[htbp]
	\vspace{-0.5cm}
	\centering
	\subfloat[Possible viewpoint zone and $\varepsilon$-viewpoint-sensitive neighborhood.]{\label{Fig:PL_prob_given_g}
		\begin{minipage}[c]{0.3\linewidth}
			\centering
			\includegraphics[width=1\textwidth]{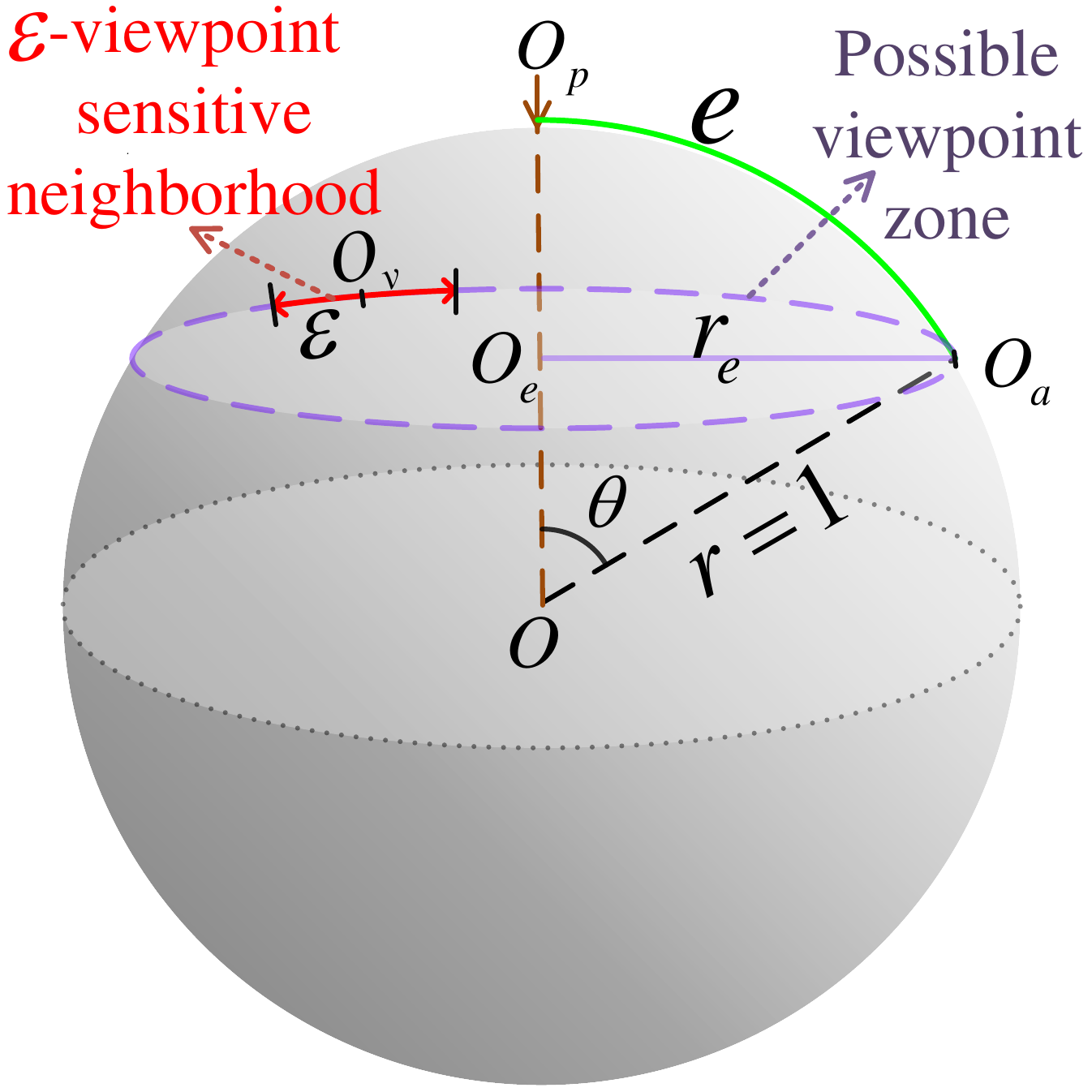}
		\end{minipage}
	}
	\subfloat[$\mathrm{Pr}_{\mathrm{e}}$ v.s. $e$ and $\varepsilon$]{\label{Fig:PL_prob_given_g_3D_numerical}
		\begin{minipage}[c]{0.35\linewidth}
			\centering
			\includegraphics[width=1\textwidth]{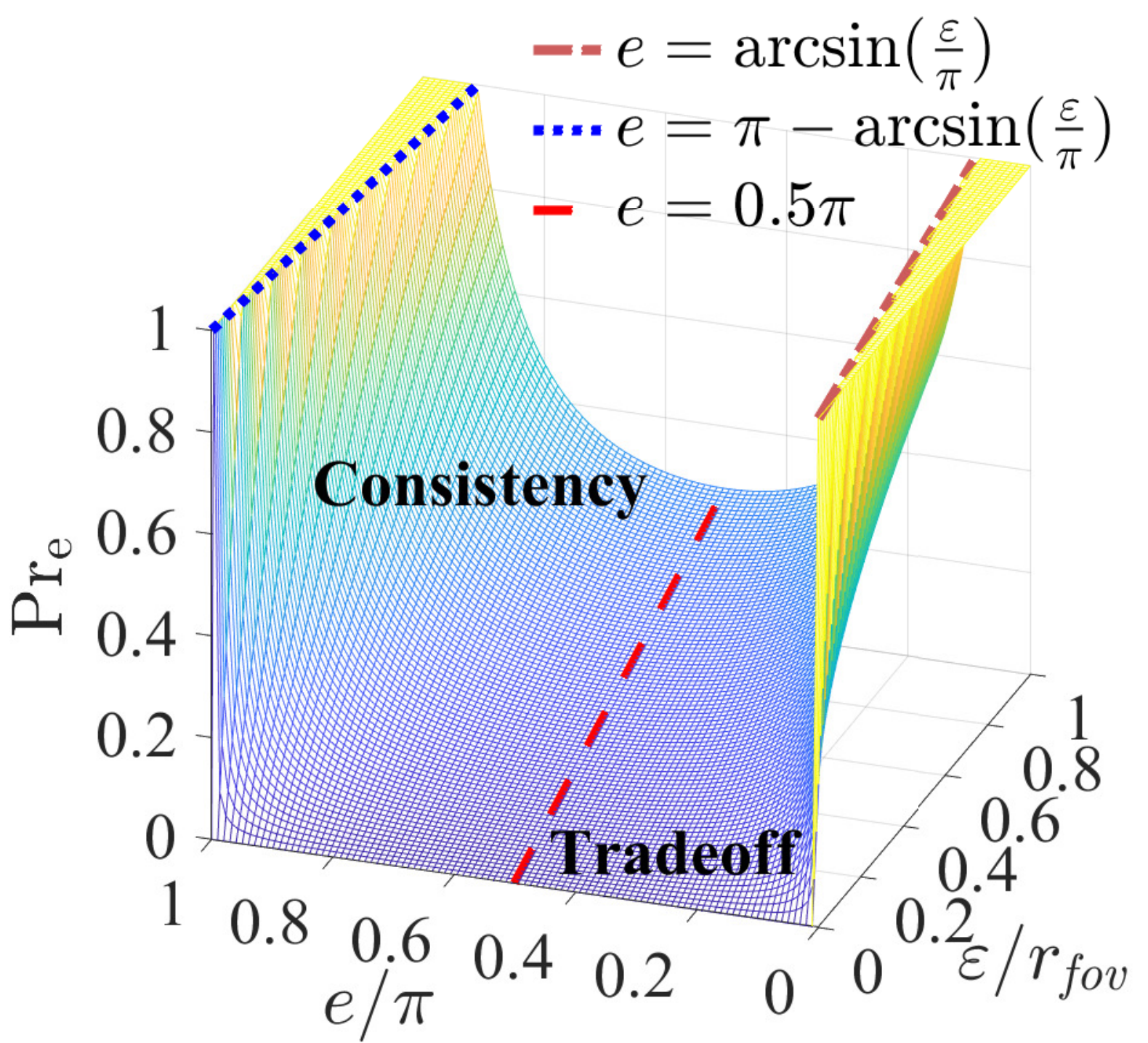}
		\end{minipage}
	}
		\subfloat[$\mathrm{Pr}_{\mathrm{e}}$ v.s. $e$, $\varepsilon=0.4r_{\textit{fov}}$.]{\label{Fig:PL_prob_given_g_2D_numerical}
		\begin{minipage}[c]{0.35\linewidth}
			\centering
			\includegraphics[width=1\textwidth]{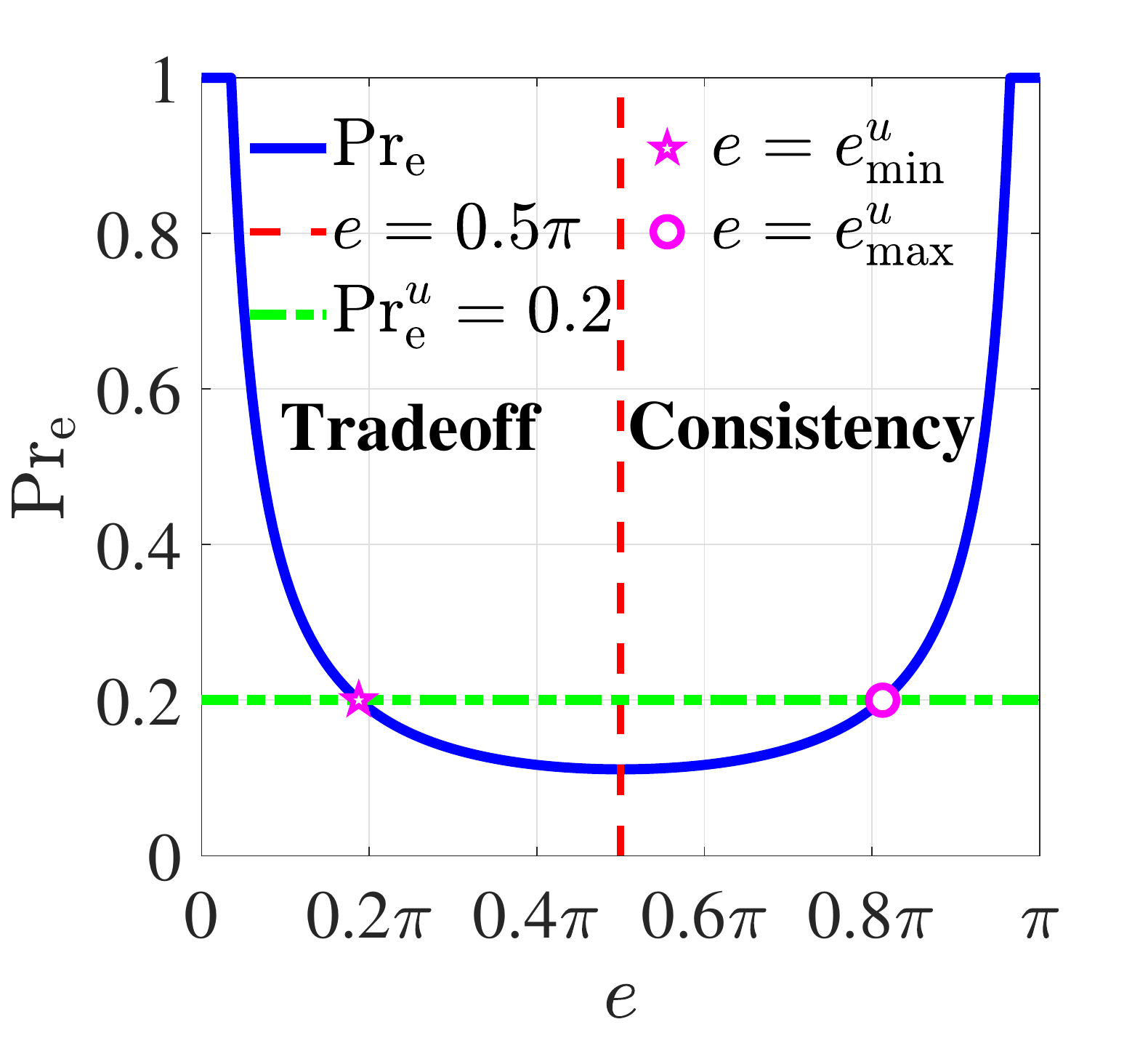}
		\end{minipage}
	}	
	\vspace{-0.35cm}
	\caption{Viewpoint leakage probability when uploading $e$. There exist a tradeoff region and a consistent region between $\mathrm{Pr}_{\mathrm{e}}$ and $\mathrm{QoE}$.}\label{Fig:PL_prob_given_g_numerical}	
	\vspace{-0.75cm}
\end{figure}

When the server obtains the predicted viewpoint $O_p$ and prediction error $e$, one can only infer that the real viewpoint $O_v$ is on a circle, i.e., the possible viewpoint zone is a circle
with center $O_e$ as shown in Fig. \ref{Fig:PL_prob_given_g}. The $\varepsilon$-viewpoint-sensitive neighborhood becomes an arc of the circle with center $O_v$.  In the sequel, we first derive the measures of the zone $\mathcal{Z}_v$ and neighborhood $\mathcal{N}(O_v, \varepsilon)$, respectively, based on which we obtain the leakage probability.

The measure of $\mathcal{Z}_v$ is the circumference of the circle, i.e., $\lambda[\mathcal{Z}_v]=2\pi r_e$,
where $r_e$ is the radius of the circle and is a function of $e$ as derived in the following.
The straight-line distance from an arbitrary viewpoint on the circle $O_a$ to the center of the sphere $O$ is the radius of the sphere $r=1$. The angle between the line $\overline{O_p O}$ and the line $\overline{O_a O}$ is $\theta$. When measured in radian, $\theta = e$. The radius of the circle is $r_e = \sin(\theta) \cdot r = \sin(e)$. By substituting $r_e=\sin(e)$ into $\lambda[\mathcal{Z}_v]=2\pi r_e$, the measure of $\mathcal{Z}_v$ is $\lambda[\mathcal{Z}_v]=2\pi \sin(e)$.

The measure of $\mathcal{N}(O_v, \varepsilon)$ is the arc length $2\varepsilon$. Since the length of the neighborhood is no more than the circumference of the circle, $\lambda[\mathcal{N}(O_v, \varepsilon)]=\min\{2\varepsilon, 2\pi \sin(e)\}$.

According to Definition \ref{def_VLP_inital}, the $\varepsilon$-viewpoint leakage probability when uploading $e$ is
\begin{align}\label{PL_prob_circle}
\mathrm{Pr}_{\mathrm{e}}\triangleq\min\left\{\frac{\min\{2\varepsilon, 2\pi \sin(e)\}}{2\pi \sin(e)},1\right\} = \min\left\{\frac{\varepsilon}{\pi\sin(e)}, 1\right\}
\end{align}
which increases as $\sin(e)$ decreases. Then, the minimal viewpoint leakage probability is $\mathrm{Pr}_{\mathrm{e}}^{\min}=\frac{\varepsilon}{\pi}$, which is achieved when $\sin(e)$ achieves the maximum, i.e., $e=0.5\pi$.

From \eqref{PL_prob_circle}, $\mathrm{Pr}_{\mathrm{e}}=1$ when $\pi \sin(e)\leq \varepsilon $, from which we obtain the range of $e$ as
\begin{align}\label{g_when_PL_prob_circle=1}
e\in[0, \arcsin(\frac{\varepsilon}{\pi})]\cup[\pi -  \arcsin(\frac{\varepsilon}{\pi}),\pi]
\end{align}
If $e$ falls in this range, $\varepsilon$-viewpoint leakage event happens, the inferred viewpoint can be arbitrary viewpoint in the possible viewpoint zone. In Fig. \ref{Fig:PL_prob_given_g_3D_numerical}, we provide the value of $\mathrm{Pr}_{\mathrm{e}}$ given the normalized values of $e$ and $\varepsilon$ computed from \eqref{PL_prob_circle}.

\vspace{-3mm}\subsection{Relations of Viewpoint Leakage with QoE and Prediction Performance }\label{subsection:upload_g_Tradeoff_g_viewpoint_privacy}

From \eqref{PL_prob_circle}, as the prediction error $e$ increases from $\arcsin(\frac{\varepsilon}{\pi})$ to $\pi -  \arcsin(\frac{\varepsilon}{\pi})$,   $\mathrm{Pr}_{\mathrm{e}}$ first decreases and then increases, and is mirror symmetric with respect to (w.r.t.) $e=0.5\pi$, as illustrated in Fig. \ref{Fig:PL_prob_given_g_3D_numerical} and \ref{Fig:PL_prob_given_g_2D_numerical}. 

When $e\in[\arcsin(\frac{\varepsilon}{\pi}), 0.5\pi]$, the viewpoint leakage probability $\mathrm{Pr}_{\mathrm{e}}$ decreases as $e$ increases, i.e., privacy can be better protected with larger prediction error, which will degrade $\mathrm{QoE}$. Then, \textit{there is a tradeoff between protecting privacy and improving QoE} in this region. When $e\in[0.5\pi, \pi - \arcsin(\frac{\varepsilon}{\pi})]$, $\mathrm{Pr}_{\mathrm{e}}$ decreases as $e$ decreases, i.e., privacy can be better protected with smaller prediction error, and improving privacy protection is consistent with improving QoE. 
In other words, to reduce the viewpoint leakage probability, the prediction error can neither be too small nor be too large, depending on the viewpoint privacy requirement of a user.

The following corollary provides the relation of required prediction error with the viewpoint privacy requirement. The detailed proof is provided in Appendix A of \cite{Xing_viewpoint_leakage_arxiv}.
\vspace{-0.5cm}
\begin{XingCorlay}\label{corollary_1}
	Denote the maximal viewpoint leakage probability allowed by a user as $\mathrm{Pr}_{\mathrm{e}}^u$, the required range of prediction error for satisfying $\mathrm{Pr}_{\mathrm{e}}\leq\mathrm{Pr}_{\mathrm{e}}^u$ is
	\begin{align}
		e\in
		\left
		\{\begin{array}{lr}
			[0,\pi],& \mathrm{Pr}_{\mathrm{e}}^u=1,\\
			\left[e_{\min}^u, e_{\max}^u\right],& \mathrm{Pr}_{\mathrm{e}}^u\in[\mathrm{Pr}_{\mathrm{e}}^{\min},1),\\
			\textrm{Infeasible}, & \mathrm{Pr}_{\mathrm{e}}^u\in[0,\mathrm{Pr}_{\mathrm{e}}^{\min}).
		\end{array}
		\right.\label{user_req_g}
	\end{align}
	where $e_{\min}^u\triangleq\arcsin(\frac{\varepsilon}{\mathrm{Pr}_{\mathrm{e}}^u\pi})$ and $e_{\max}^u\triangleq\pi - \arcsin(\frac{\varepsilon}{\mathrm{Pr}_{\mathrm{e}}^u\pi})$.
\end{XingCorlay}
\vspace{-0.5cm}

The corollary implies that if a user has no privacy viewpoint requirement, i.e., $ \mathrm{Pr}_{\mathrm{e}}^u=1$, then the prediction error should be minimized for maximizing the QoE. When $\mathrm{Pr}_{\mathrm{e}}^u\in[\mathrm{Pr}_{\mathrm{e}}^{\min},1)$,  \textit{there is a contradiction between the target of satisfying the viewpoint privacy requirement and the target of maximizing the QoE.}
For example, in Fig. \ref{Fig:PL_prob_given_g_2D_numerical}, when $\mathrm{Pr}_{\mathrm{e}}^u=0.2$ and $\varepsilon=0.4r_{\textit{fov}}$, to satisfy the privacy requirement, $e\geq e_{\min}^u=0.19\pi$, while to maximize QoE, $e$ should be minimized.
\vspace{-0.5cm}
\begin{XingRmk}\label{Remark:uploading_e_tradeoff}
When $e\in\left[e_{\min}^u, e_{\max}^u\right]$, the privacy-QoE tradeoff exists if $e\in\left[e_{\min}^u, 0.5\pi\right]$, otherwise protecting privacy is consistent with improving QoE.    
\end{XingRmk}
\vspace{-0.5cm}

\vspace{-0.2cm}
\section{Viewpoint Leakage Probability when Uploading the QoE Metric}
\vspace{-0.1cm}
Although the real viewpoint $O_v$ cannot be inferred directly from the QoE metric in \eqref{QoE_def}, the prediction error $e$ can be inferred from the QoE.
In this section, we first derive $\mathrm{QoE}$ as a function of $r_{\textit{sv}}$ (which reflects the amount of configured resources as stated in Remark \ref{Remark:r_sv_is_resources}) and $e$, from which the prediction error can be inferred.
Then, we derive the $\varepsilon$-viewpoint leakage probability when uploading the QoE metric.

\vspace{-0.2cm}\subsection{QoE as a Function of Configured Resources and Prediction Error}\label{section:QoE}
If $r_{\textit{sv}} = 0$ (i.e., no streaming), then $\mathrm{QoE} = 0$. If $r_{\textit{sv}}=\pi$ (i.e., streaming the sphere), then $\mathrm{QoE} =100\%$.
The QoE metric as a function of $r_{\textit{sv}}\in(0,\pi)$ and $e$ in different cases is provided in the following proposition. The detailed proof is provided in Appendix B of \cite{Xing_viewpoint_leakage_arxiv}.
\vspace{-0.5cm}
\begin{XingPropo}\label{Proposition:QoE_expression}
	When $r_{\textit{sv}}\in(0,\pi)$, the expressions of $\mathrm{QoE}$ w.r.t. $r_{\textit{sv}}$ and $e$ are as follows.
	\begin{subequations}\label{QoE_funct}
	\begin{align}
			&\mathrm{QoE} = 100\%, \textrm{if} \ r_{\textit{sv}}\geq r_{\textit{fov}} + e ~ (\textrm{i.e., }\textit{FoV \!$\subset$\! SFoV}).\label{QoE_funct_fov_in_sfov}\\
			&\mathrm{QoE} = \frac{1 - \cos(r_{\textit{sv}})}{1 - \cos(r_{\textit{fov}})}, \textrm{if} \ r_{\textit{fov}}\geq r_{\textit{sv}} + e ~ (\textrm{i.e., }
			\textit{SFoV \!$\subset$\! FoV}).\\
			&\mathrm{QoE} = 0, \textrm{if} \ e\geq r_{\textit{fov}} + r_{\textit{sv}}  (\textrm{i.e., }\textit{FoV \!$\cap$\! SFoV}=\emptyset).\\
			&\mathrm{QoE} = \frac{- \cos(r_{\textit{sv}}) - \cos(r_{\textit{fov}})}{1 - \cos(r_{\textit{fov}})}, \textrm{if} \ r_{\textit{fov}} + r_{\textit{sv}} + e\geq 2\pi ~(\textrm{i.e., }\textit{SFoV-C$\subset$FoV}).\\	
			&\mathrm{QoE} =\frac{A_{\textit{ol}}^{rm}(r_{\textit{sv}},e)}{2\pi (1 - \cos (r_{\textit{fov}}))}, \textrm{if} \ r_{\textit{sv}}\in\big[r_{\textit{sv},\min}^{rm}(e), r_{\textit{sv},\max}^{rm}(e)\big] ~(\textrm{i.e., }\textrm{remaining case}),  \label{QoE_funct_remaining_case}
	\end{align}
\end{subequations}
where
\begin{align}\label{S_overlap}
	A_{\textit{ol}}^{rm}(r_{\textit{sv}},e)\triangleq& 2\pi - 2\pi\cos (r_{\textit{sv}}) - 2\pi\cos (r_{\textit{fov}}) - 2\arccos \left(\frac{\cos (e) - \cos(r_{\textit{sv}})\cos(r_{\textit{fov}})}{\sin(r_{\textit{sv}})\sin(r_{\textit{fov}})}\right) \nonumber\\
	& + 2\cos(r_{\textit{sv}}) \arccos \left(\frac{-\cos(r_{\textit{fov}}) + \cos (e) \cos(r_{\textit{sv}})}{\sin(e)\sin(r_{\textit{sv}})}\right) \nonumber\\
	& + 2\cos(r_{\textit{fov}}) \arccos\left( \frac{-\cos(r_{\textit{sv}}) + \cos (e)\cos(r_{\textit{fov}})}{\sin(e)\sin(r_{\textit{fov}})} \right),
\end{align}
$r_{\textit{sv},\min}^{rm}(e)\triangleq |r_{\textit{fov}} - e|$, and $r_{\textit{sv},\max}^{rm}(e)\triangleq \min\{r_{\textit{fov}} + e, 2\pi - (r_{\textit{fov}} + e) \}$.
\end{XingPropo}
\vspace{-0.5cm}

\begin{figure}[htbp]
	\vspace{-1.35cm}
	\centering
	\subfloat[\textit{FoV \!$\subset$\! SFoV}]{\label{Fig:r_fov_r_str_g_fov_in_cc}
		\begin{minipage}[c]{0.33\linewidth}
			\centering
			\includegraphics[width=1\textwidth]{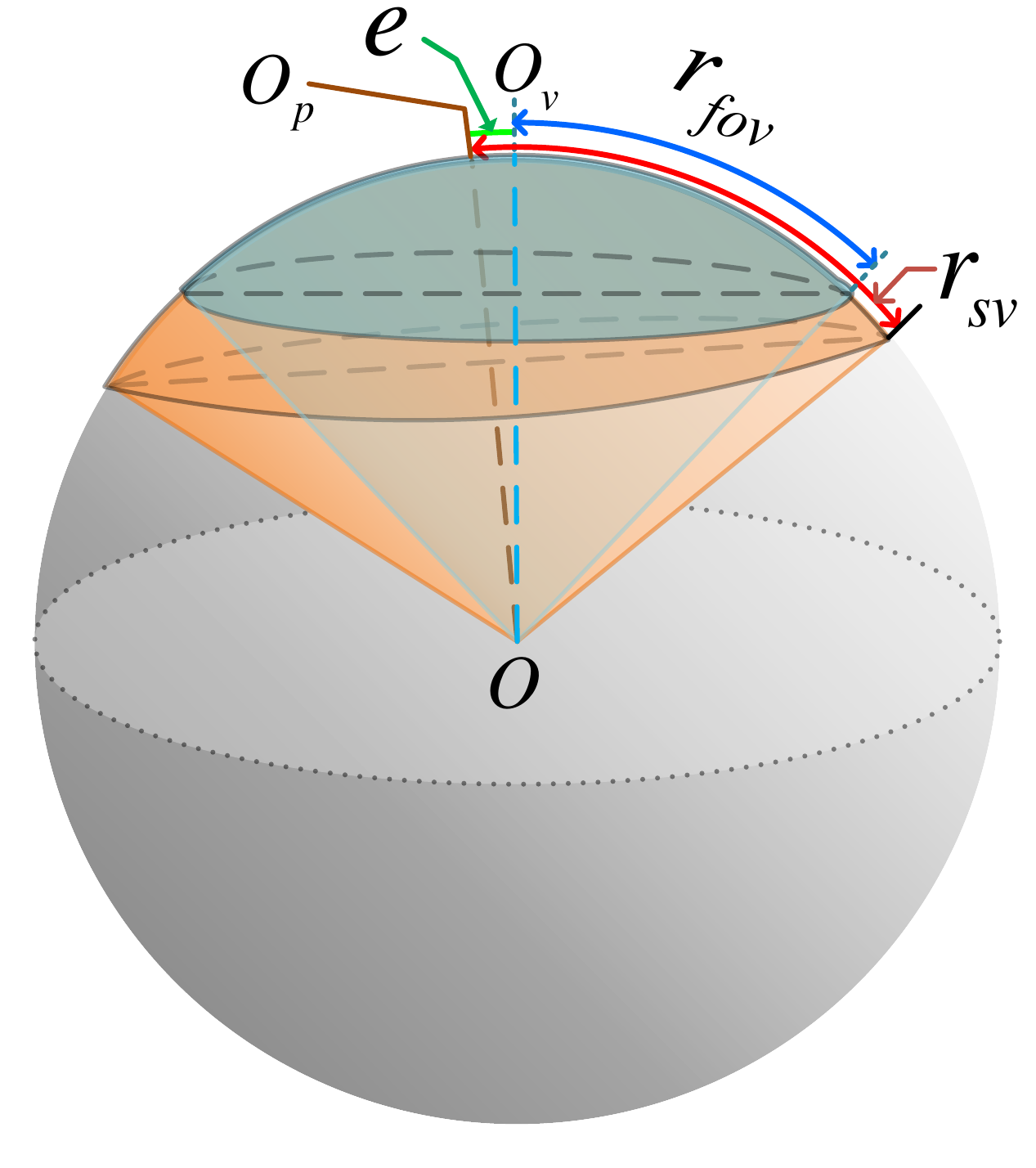}
		\end{minipage}
	}
	\subfloat[\textit{SFoV \!$\subset$\! FoV}]{\label{Fig:r_fov_r_str_g_cc_in_fov}
		\begin{minipage}[c]{0.33\linewidth}
			\centering
			\includegraphics[width=1\textwidth]{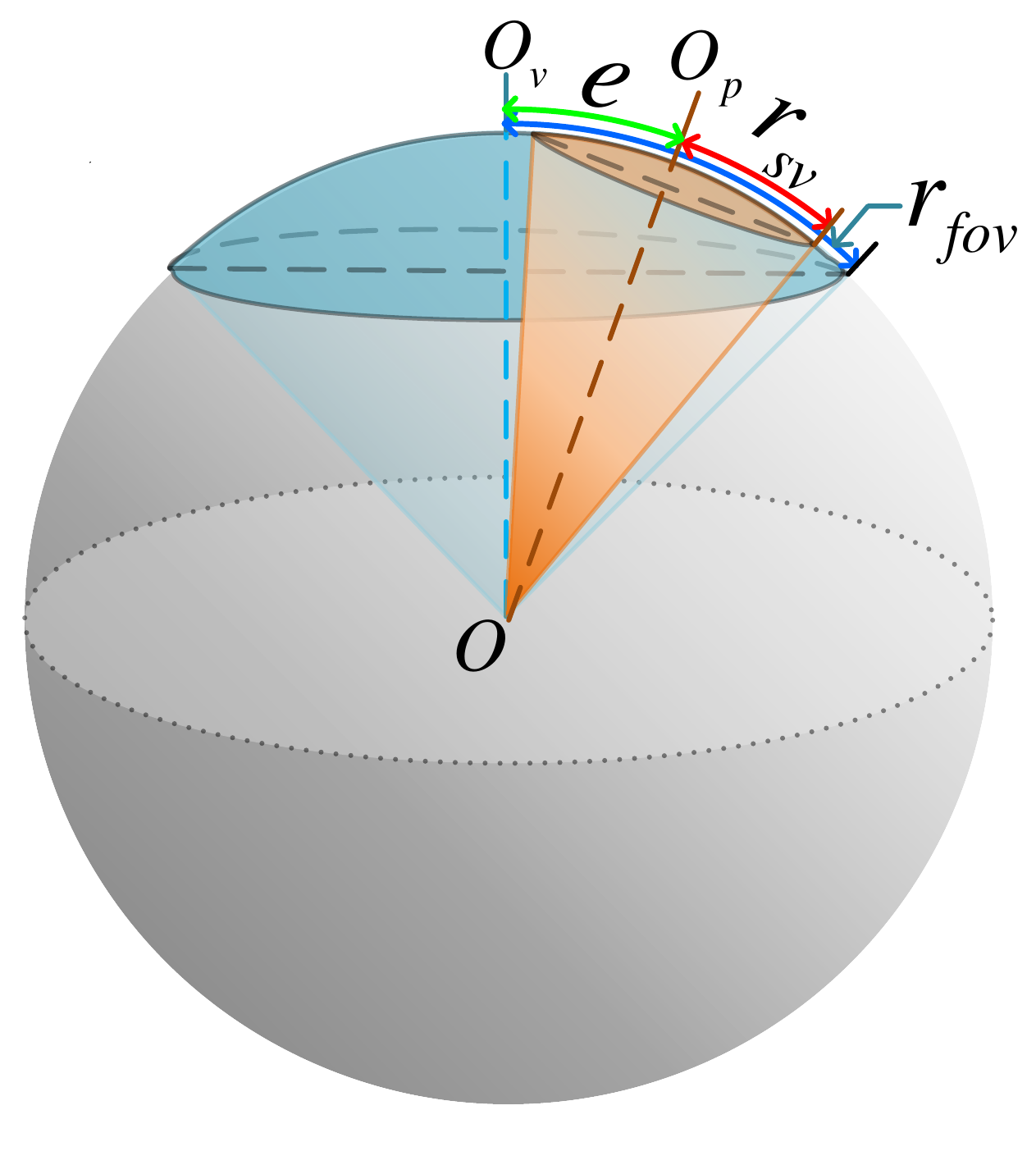}
		\end{minipage}
	}
	\subfloat[\textit{FoV \!$\cap$\! SFoV} = $\emptyset$]{\label{Fig:r_fov_r_str_g_fov_cc_no_intersection}
		\begin{minipage}[c]{0.33\linewidth}
			\centering
			\includegraphics[width=1\textwidth]{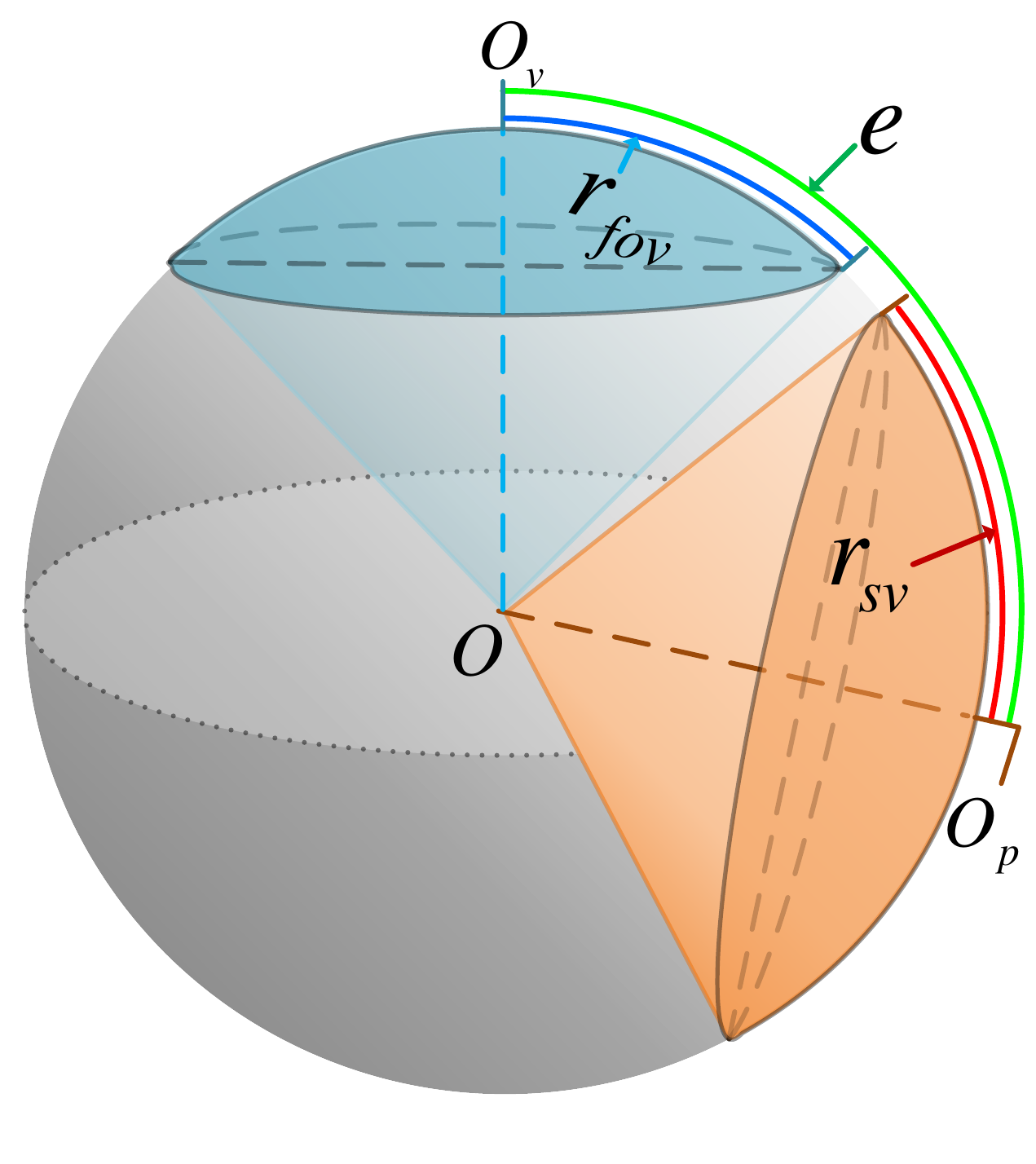}
		\end{minipage}
	}\\
	\subfloat[\textit{SFoV-C$\subset$FoV}]{\label{Fig:r_fov_r_str_g_inverse_cc_in_fov}
		\begin{minipage}[c]{0.33\linewidth}
			\centering
			\includegraphics[width=1\textwidth]{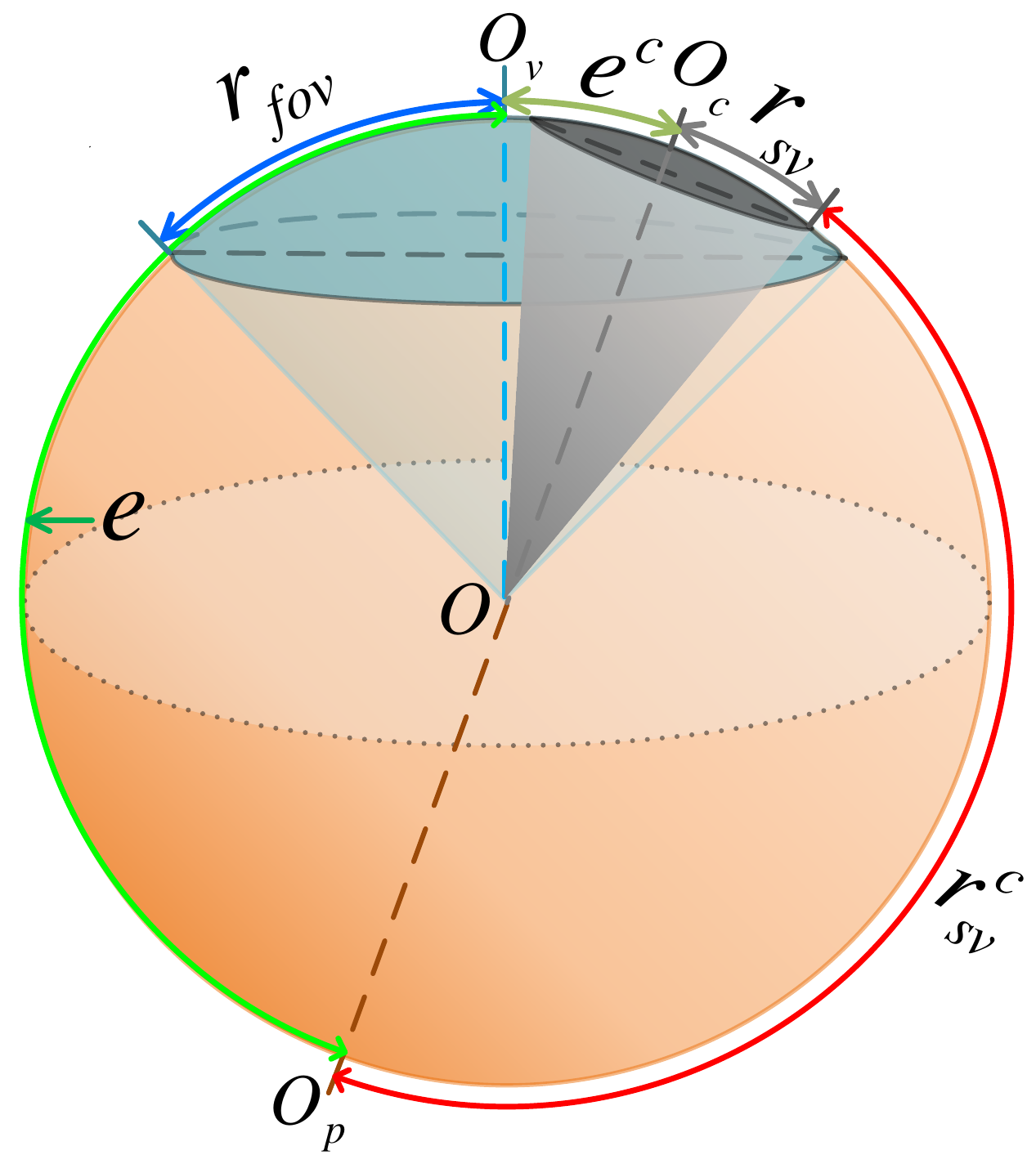}
		\end{minipage}
	}
		\subfloat[QoE v.s. $ r_{\textit{sv}}$ and $e$.]{\label{Fig:qoe_r_g_relation}
		\begin{minipage}[c]{0.5\linewidth}
			\centering
			\includegraphics[width=1\textwidth]{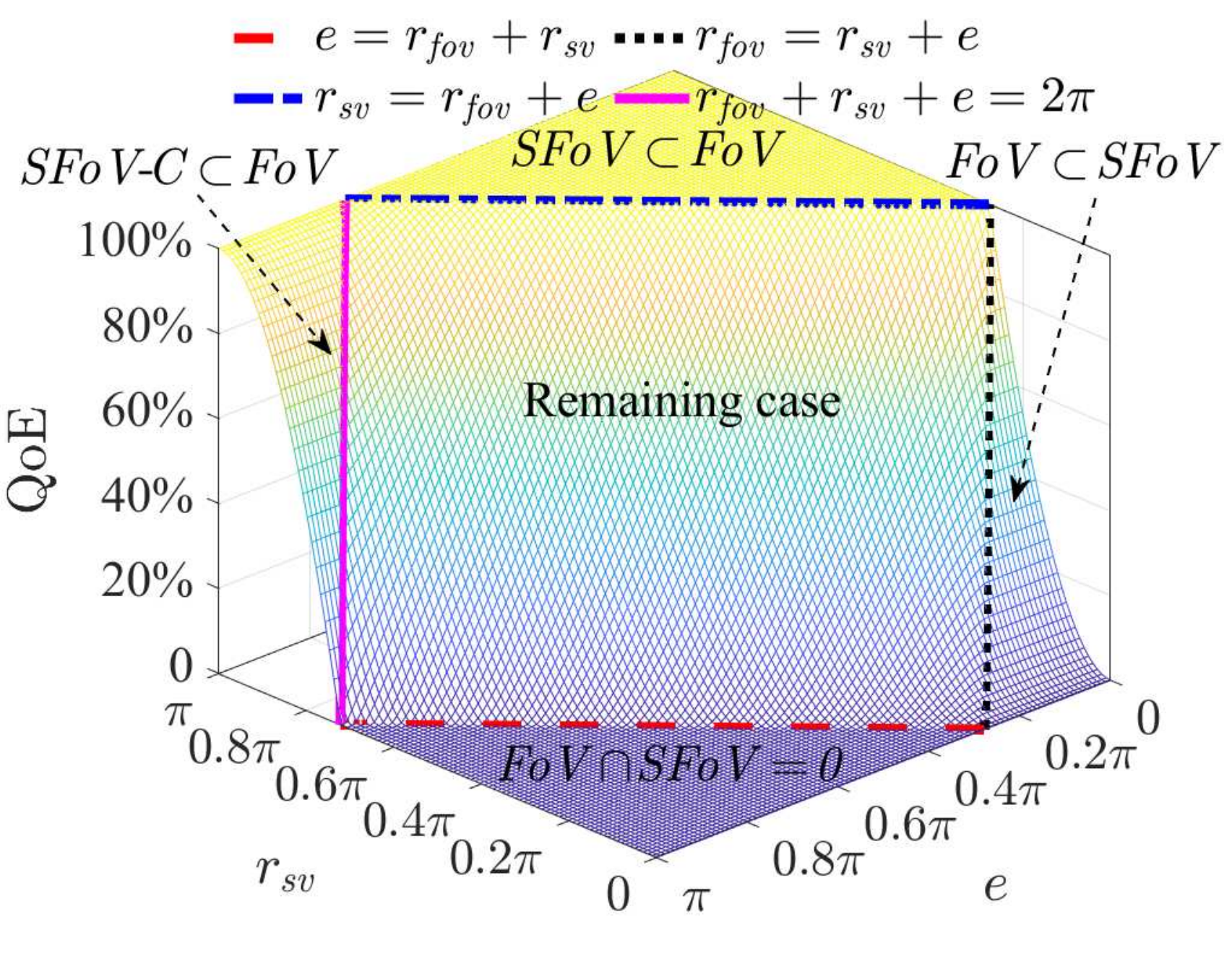}
		\end{minipage}
	}
	\vspace{-0.45cm}
	\caption{Four cases when $r_{\textit{sv}}\in(0,\pi)$, and $\mathrm{QoE}$ v.s. $r_{\textit{sv}}$ and $e$.}\label{Fig:r_fov_r_str_g_all}
	\vspace{-1.3cm}	
\end{figure}

In the proposition, \textit{FoV \!$\subset$\! SFoV}, \textit{SFoV \!$\subset$\! FoV}, \textit{FoV \!$\cap$\! SFoV}=$\emptyset$, and \textit{SFoV-C$\subset$FoV} respectively denote the cases that FoV being included in SFoV, SFoV being included in FoV, no intersection between FoV and SFoV, and the complement set of SFoV being included in FoV. These four cases are illustrated in Fig. \eqref{Fig:r_fov_r_str_g_fov_in_cc}-\eqref{Fig:r_fov_r_str_g_inverse_cc_in_fov}.
\vspace{-0.5cm}
\begin{XingRmk}\label{Remark:relation_qoe_four_cases}
	In the four cases that \textit{FoV \!$\subset$\! SFoV}, \textit{SFoV \!$\subset$\! FoV}, \textit{FoV \!$\cap$\! SFoV}=$\emptyset$, or \textit{SFoV-C$\subset$FoV}, $\mathrm{QoE}$ does not depend on prediction error $e$.	
\end{XingRmk}
\vspace{-0.5cm}
\vspace{-0.5cm}
\begin{XingRmk}\label{Remark:relation_qoe_five_case}
	In the remaining case, $\mathrm{QoE}=\frac{A_{\textit{ol}}^{rm}(r_{\textit{sv}},e)}{2\pi (1 - \cos (r_{\textit{fov}}))}$, which is a strictly monotonically decreasing function of $e$ and increasing function of $r_{\textit{sv}}$.
\end{XingRmk}
\vspace{-0.5cm}

To visualize the impact of configured resources $r_{\textit{sv}}$ and prediction error $e$ on $\mathrm{QoE}$ as stated in Remarks 3 and 4, we provide the values of $\mathrm{QoE}$ obtained from \eqref{QoE_funct} given all possible values of $r_{\textit{sv}}$ and $e$ in Fig. \ref{Fig:qoe_r_g_relation}.

\subsection{Inferring Prediction Error from $\mathrm{\mathrm{QoE}}$}
Except the QoE metric, the server also knows the predicted viewpoint $O_p$, the radius of the FoV $r_{\textit{fov}}$, and the radius of the SFoV $r_{\textit{sv}}$ for proactive VR video streaming.

The following proposition provides the prediction error inferred from these information.
The proof is provided in Appendix C of \cite{Xing_viewpoint_leakage_arxiv}.
\vspace{-0.5cm}
\begin{XingPropo}\label{Proposition:infer_case_from_QoE}
	When $r_{\textit{sv}}\in(0,\pi)$, given the value of $\mathrm{QoE}$, the range or value of $e$ is
	\vspace{-0.3cm}
	\begin{subequations}
	\begin{align}
			&e\in[0, r_{\textit{sv}} - r_{\textit{fov}}], \textrm{if} \ \mathrm{QoE}=100\%\ (\textrm{i.e., }\textit{FoV \!$\subset$\! SFoV}). \label{corlay_expression_1}\\
			&e\in[0, r_{\textit{fov}} - r_{\textit{sv}}], \textrm{if} \ \mathrm{QoE} = \frac{1 - \cos(r_{\textit{sv}})}{1 - \cos(r_{\textit{fov}})}\ (\textrm{i.e., }\textit{SFoV \!$\subset$\! FoV}). \label{corlay_expression_2}\\
			&e\in[r_{\textit{fov}} + r_{\textit{sv}},\pi], \textrm{if} \ \mathrm{QoE} = 0\ (\textrm{i.e., }  \textit{FoV \!$\cap$\! SFoV}=\emptyset).\label{corlay_expression_3}\\
			&e\in[2\pi - r_{\textit{fov}} - r_{\textit{sv}},\pi], \textrm{if} \ \mathrm{QoE}= \frac{- \cos(r_{\textit{sv}}) - \cos(r_{\textit{fov}})}{1 - \cos(r_{\textit{fov}})} (\textrm{i.e., }  \textit{SFoV-C$\subset$FoV}). \label{corlay_expression_4}\\
			&e=e^{bi}, \textrm{otherwise (i.e., remaining case)}. \label{corlay_expression_5}
	\end{align}
	\end{subequations}
where $e^{bi}$ is the value of $e$ determined by \eqref{QoE_funct_remaining_case} with bisection searching.
\end{XingPropo}
\vspace{-0.7cm}

\subsection{$\varepsilon$-Viewpoint Leakage Probability}
Based on Propositions \ref{Proposition:QoE_expression} and \ref{Proposition:infer_case_from_QoE} as well as Remark \ref{Remark:relation_qoe_five_case}, we can derive the $\varepsilon$-viewpoint leakage probability when uploading the QoE metric, denoted as $\mathrm{Pr}_{\mathrm{q}}$.
When $r_{\textit{sv}}=0$ or $\pi$, neither the range nor the value of $e$ can be inferred. Then, the leakage probability achieves its minimum as $\mathrm{Pr}_{\mathrm{q}}^{\min} = \mathrm{Pr}_{}^{\min} = \frac{1 - \cos(\varepsilon)}{2}$.

In the following, we derive the probability when $r_{\textit{sv}}\in(0,\pi)$.

\begin{figure}[htbp]
	\centering
	\subfloat[\textit{FoV}$\subset$\textit{SFoV} or \textit{SFoV}$\subset$\textit{FoV}]{\label{Fig:Viewpoint_concealed_zone_first_two}
		\begin{minipage}[c]{0.5\linewidth}
			\centering
			\includegraphics[width=1\textwidth]{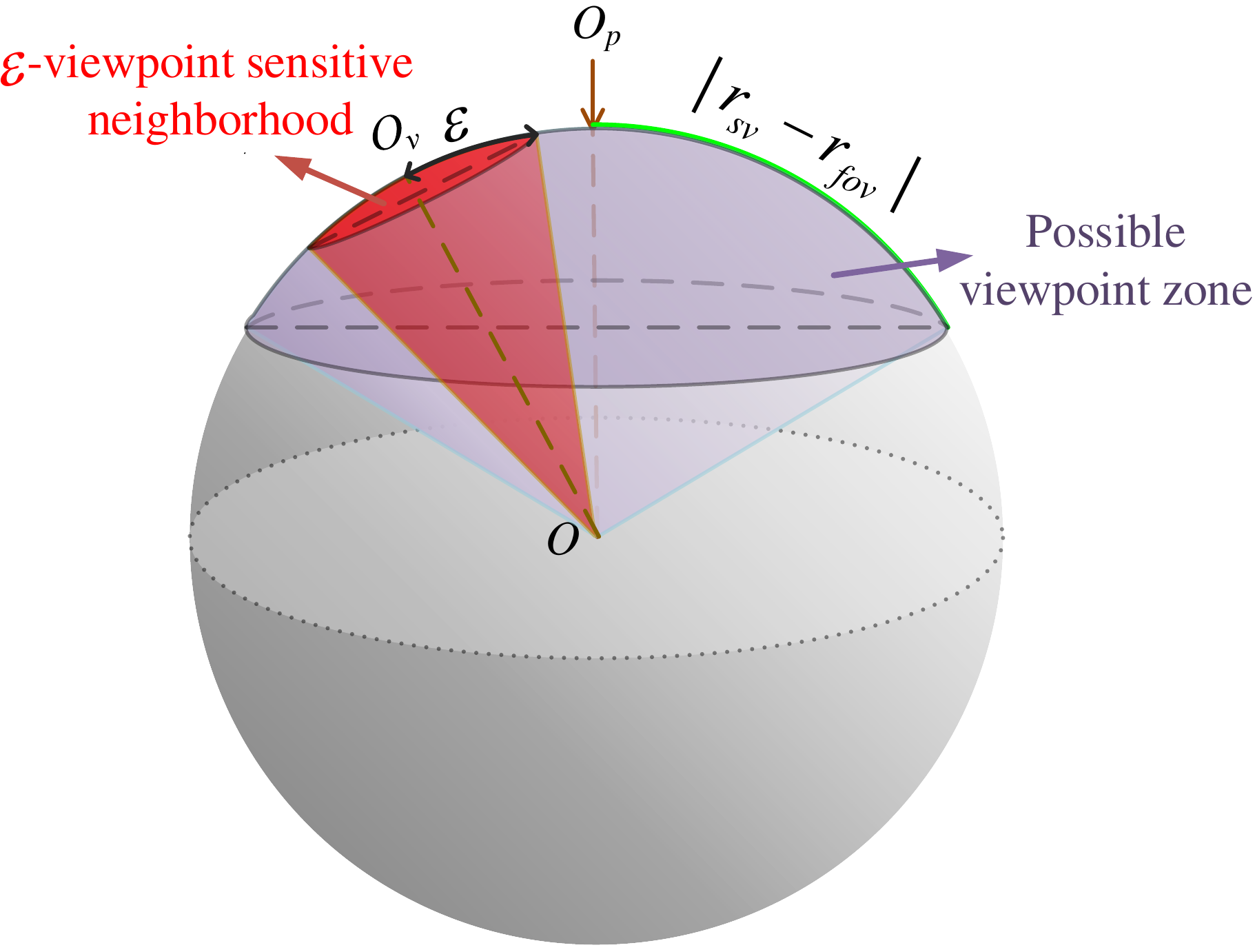}
		\end{minipage}
	}
	\subfloat[\textit{FoV \!$\cap$\! SFoV}=$\emptyset$ or \textit{SFoV-C$\subset$FoV} ]{\label{Fig:Viewpoint_concealed_zone_fov_cc_next_two}
		\begin{minipage}[c]{0.5\linewidth}
			\centering
			\includegraphics[width=1\textwidth]{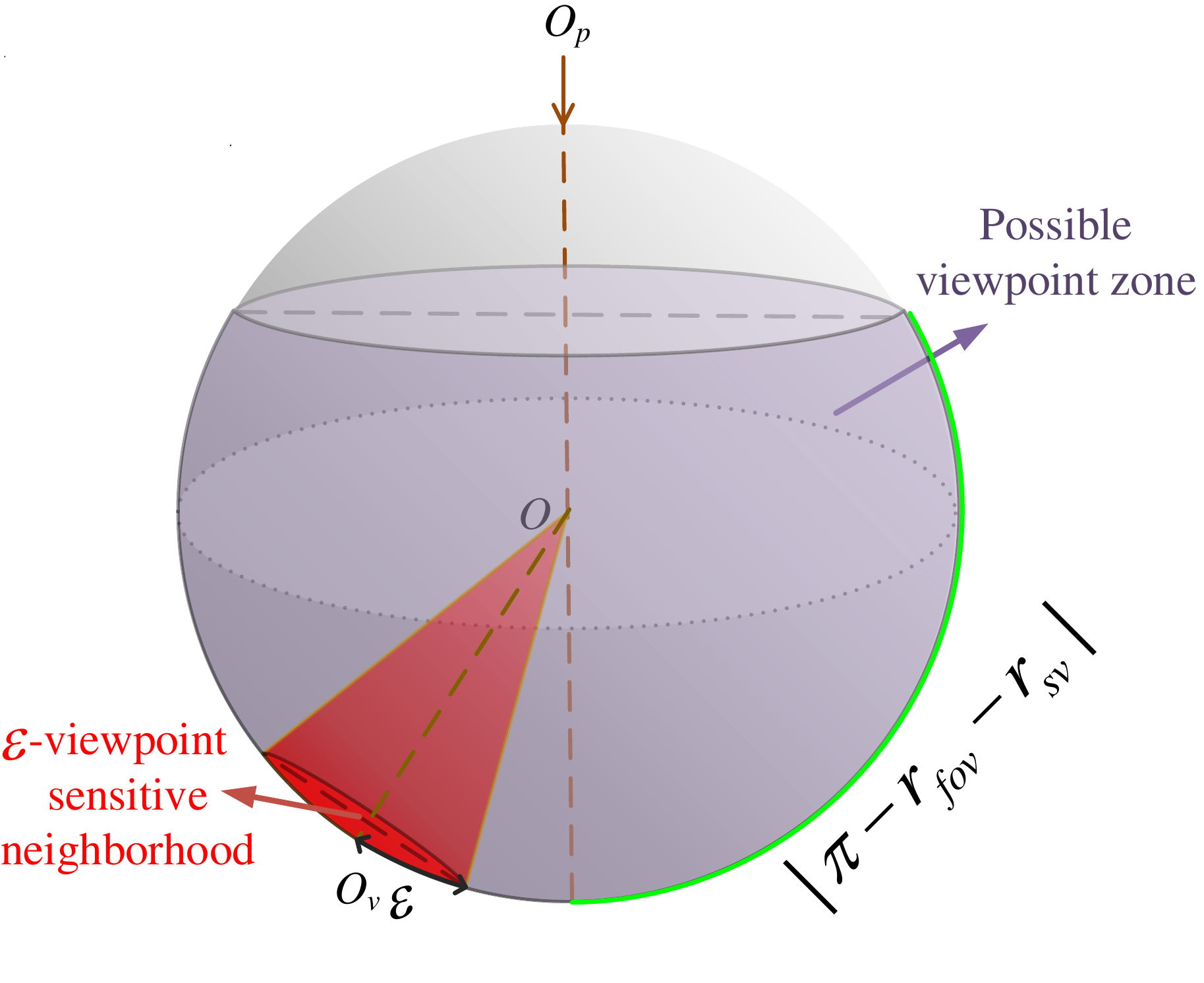}
		\end{minipage}
	}
	\vspace{-0.35cm}
	\caption{Possible viewpoint zone when \textit{FoV}$\subset$\textit{SFoV}, \textit{SFoV}$\subset$\textit{FoV}, \textit{FoV \!$\cap$\! SFoV}=$\emptyset$, and \textit{SFoV-C$\subset$FoV} }\label{Fig:Viewpoint_concealed_zone_four_regions}	
	\vspace{-0.95cm}
\end{figure}

\subsubsection{\textit{FoV}$\subset$\textit{SFoV}}\label{subsubsection:PL_F_in_S}
From \eqref{corlay_expression_1}, $ e\in[0, r_{\textit{sv}} - r_{\textit{fov}}]$. As shown in Fig. \ref{Fig:Viewpoint_concealed_zone_first_two},
given the predicted viewpoint $O_p$, the possible viewpoint zone $\mathcal{Z}_v$ is a spherical cap with radius $r_z = r_{\textit{sv}} - r_{\textit{fov}}\geq0$ and area $A_{\mathcal{Z}_v}=2\pi(1 - \cos(r_{z}))$. The $\varepsilon$-viewpoint-sensitive neighborhood is a spherical cap with center $O_v$ and radius $\varepsilon$, whose area is $A_{\mathcal{N}(O_v,\varepsilon)}=2\pi(1 - \cos(\varepsilon))$. Then, according to Definition \ref{def_VLP_inital}, the $\varepsilon$-viewpoint leakage probability is
\begin{align*}
\mathrm{Pr}_{\mathrm{q}}^{\textit{V}\subset \textit{SV}}(\varepsilon,r_{\textit{sv}}) = \min\left\{\frac{2\pi(1-\cos(\varepsilon))}{2\pi(1 - \cos(r_z))}, 1\right\}=  \min\left\{\frac{1-\cos(\varepsilon)}{1 - \cos(r_{\textit{sv}} - r_{\textit{fov}})}, 1\right\}
\end{align*}

\subsubsection{\textit{SFoV}$\subset$\textit{FoV}} From \eqref{corlay_expression_2}, $ e\in[0, r_{\textit{fov}} - r_{\textit{sv}}]$. As shown in Fig. \ref{Fig:Viewpoint_concealed_zone_first_two}, the only difference from the case where \textit{FoV}$\subset$\textit{SFoV} is that the radius of the possible viewpoint zone  becomes $r_z = r_{\textit{fov}} - r_{\textit{sv}}\geq0$. 
Then, the $\varepsilon$-viewpoint leakage probability is
\begin{align*}
	\mathrm{Pr}_{\mathrm{q}}^{\textit{SV}\subset \textit{V}}(\varepsilon,r_{\textit{sv}}) = \min\left\{\frac{1-\cos(\varepsilon)}{1 - \cos(r_{\textit{fov}} - r_{\textit{sv}})}, 1\right\}
\end{align*}

\subsubsection{\textit{FoV \!$\cap$\! SFoV}=$\emptyset$} From \eqref{corlay_expression_3}, $e\in[r_{\textit{fov}} + r_{\textit{sv}},\pi]$. As shown in Fig. \ref{Fig:Viewpoint_concealed_zone_fov_cc_next_two}, $\mathcal{Z}_v$ is a spherical cap with radius $r_z = \pi - (r_{\textit{sv}} + r_{\textit{fov}}) \geq0$.
Then, the $\varepsilon$-viewpoint leakage probability is
\begin{align*}
	\mathrm{Pr}_{\mathrm{q}}^{\textit{V} \cap \textit{SV}\!=\!\emptyset}(\varepsilon,r_{\textit{sv}})\!=\! \min\!\left\{\frac{1-\cos(\varepsilon)}{1 - \cos(\pi - (r_{\textit{sv}} + r_{\textit{fov}}))}, 1\!\right\}
	= \min\!\left\{\frac{1-\cos(\varepsilon)}{1 + \cos(r_{\textit{sv}} + r_{\textit{fov}})}, 1\!\right\}
\end{align*}

\subsubsection{\textit{SFoV-C$\subset$FoV}} \label{subsubsection:S_c_in_F}
From \eqref{corlay_expression_4}, $e\in[2\pi - r_{\textit{fov}} - r_{\textit{sv}},\pi]$. As shown in Fig. \ref{Fig:Viewpoint_concealed_zone_fov_cc_next_two}, the only difference from the case that \textit{FoV \!$\cap$\! SFoV}=$\emptyset$  is that the radius of the possible viewpoint zone becomes $r_z = \pi - (2\pi - r_{\textit{fov}} - r_{\textit{sv}}) = r_{\textit{fov}} + r_{\textit{sv}} -\pi\geq0$. The $\varepsilon$-viewpoint leakage probability is
\begin{align*}
	\mathrm{Pr}_{\mathrm{q}}^{\textit{SV-C}\subset\textit{V}}(\varepsilon,r_{\textit{sv}})= \!\min\!\left\{\frac{1-\cos(\varepsilon)}{1 - \cos(r_{\textit{fov}} + r_{\textit{sv}} -\pi)}, 1\right\} = \!\min\!\left\{\frac{1-\cos(\varepsilon)}{1 + \cos(r_{\textit{sv}} + r_{\textit{fov}})}, 1\right\} 
\end{align*}

The viewpoint leakage probability in the four cases can be unified as
\begin{align}\label{PL_prob_4_cases}
\mathrm{Pr}_{\mathrm{q}}(\varepsilon,r_{\textit{sv}})=\min\left\{\frac{1-\cos(\varepsilon)}{1 - \cos(r_z)}, 1\!\right\}
\end{align}
where $r_z=|r_{\textit{sv}} - r_{\textit{fov}}|$ for \textit{FoV}$\subset$\textit{SFoV} and \textit{SFoV}$\subset$\textit{FoV}, and $r_z=|\pi - r_{\textit{sv}} - r_{\textit{fov}}|$ for \textit{FoV \!$\cap$\! SFoV}=$\emptyset$ and \textit{SFoV-C$\subset$FoV}. With $r_z$, the corresponding range of $e$ can be expressed as follows:
\begin{align}\label{e_4_cases}
e \leq r_z, &~~~\textrm{if}\ \textit{FoV}\!\subset\!\textit{SFoV}\  \textrm{or} \ \textit{SFoV}\!\subset\!\textit{FoV} \nonumber \\\ \
e \geq \pi - r_z, &~~~\textrm{if}\ \textit{FoV} \!\cap\! \textit{SFoV}\!=\!\emptyset\ \textrm{or} \ \textit{SFoV-C}\!\subset\! \textit{FoV}
\end{align}

\subsubsection{Remaining case}
In this case, since the value of $e$ can be obtained from \eqref{QoE_funct_remaining_case} with bisection searching according to Remark \ref{Remark:relation_qoe_five_case}, $\mathrm{Pr}_{\mathrm{q}}(\varepsilon,r_{\textit{sv}})=\mathrm{Pr}_{\mathrm{e}}$. According to \eqref{PL_prob_circle}, the $\varepsilon$-viewpoint leakage probability is $\mathrm{Pr}_{\mathrm{q}}^{\textit{RM}} =\min\left\{\frac{\varepsilon}{\pi\sin(e)}, 1\right\}$.

\vspace{-0.2cm}
\begin{XingRmk}\label{Remark:VLP_e_r_sv_relation}
For the four cases, it is shown from \eqref{PL_prob_4_cases} that the viewpoint leakage probability does not depend on $e$. For the remaining case, the leakage probability does not depend on $r_{\textit{sv}}$.
\end{XingRmk}
\vspace{-0.5cm}

\begin{figure}[htbp]
	\vspace{-0.3cm}
	\centering
	\subfloat[Frontal view.]{\label{Fig:PL_prob_all_cases_front}
		\begin{minipage}[c]{0.5\linewidth}
			\centering
			\includegraphics[width=1\textwidth]{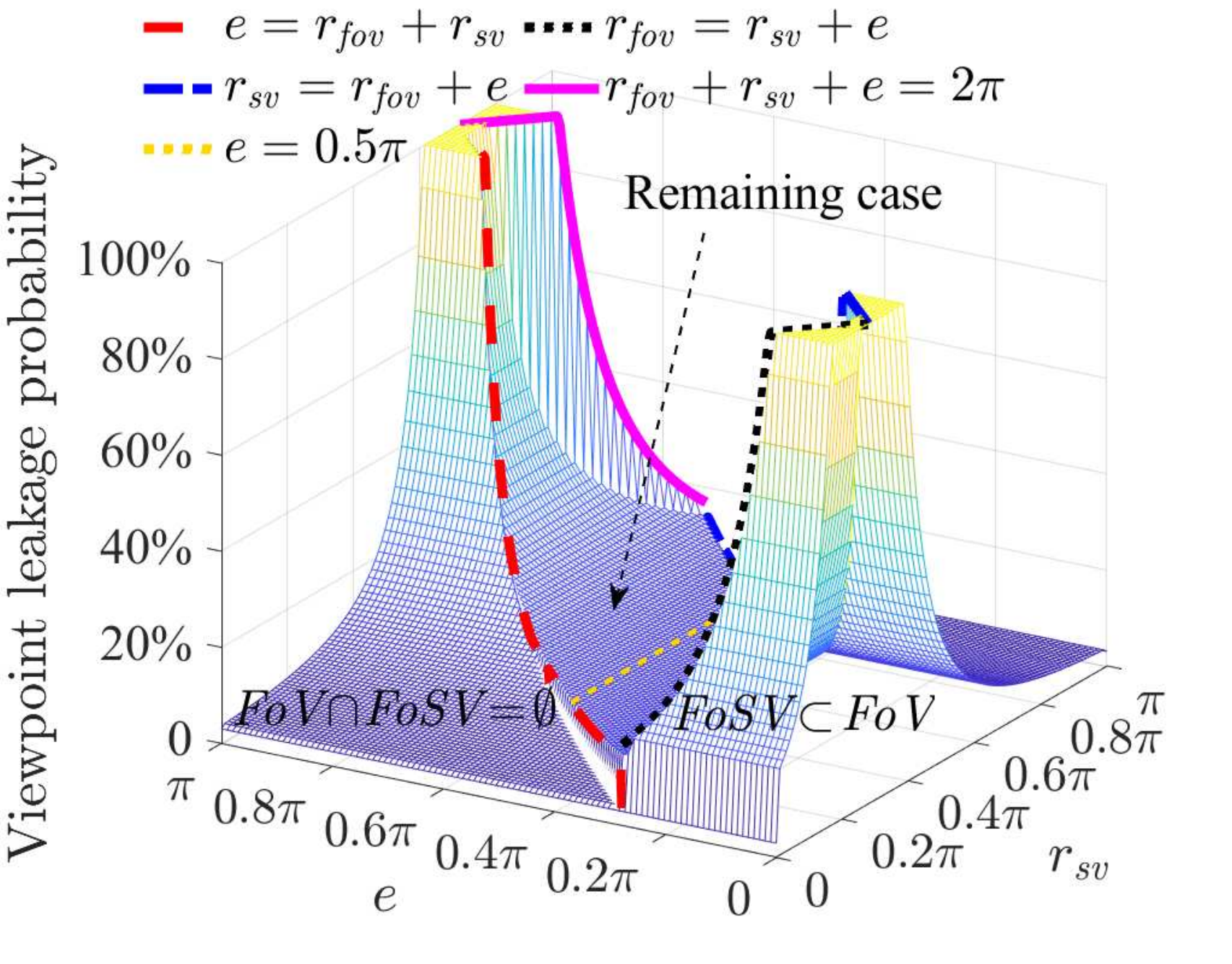}
		\end{minipage}
	}
	\subfloat[Dorsal view.]{\label{Fig:PL_prob_all_cases_back}
		\begin{minipage}[c]{0.5\linewidth}
			\centering
			\includegraphics[width=1\textwidth]{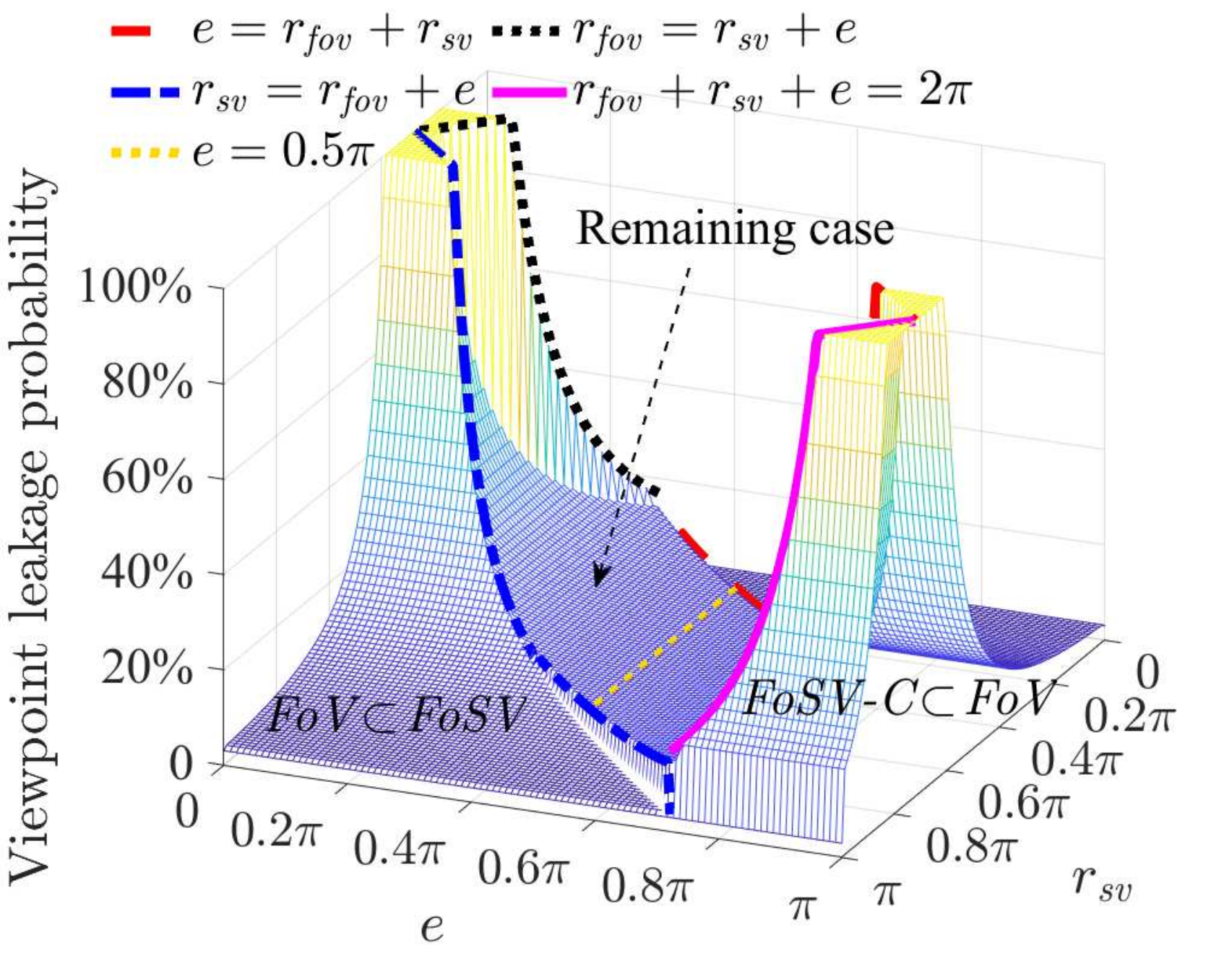}
		\end{minipage}
	}
	\vspace{-0.35cm}
	\caption{Viewpoint leakage probability v.s. $r_{\textit{sv}}$ and $e$, $\varepsilon=0.4 r_{\textit{fov}}$.}\label{Fig:PL_prob_all_cases}	
	\vspace{-0.45cm}
\end{figure}

The viewpoint leakage probability are illustrated in Fig. \ref{Fig:PL_prob_all_cases}.

\vspace{-0.6cm}
\subsection{When the $\varepsilon$-Viewpoint Leakage Probability Achieves the Maximum and Minimum?}
Since for the cases where $r_{\textit{sv}}=0$ or $\pi$, the global minimal value of $\mathrm{Pr}_{\mathrm{q}}$ can be achieved as $\mathrm{Pr}_{\mathrm{q}}^{\min} = \frac{1 - \cos(\varepsilon)}{2}$, we consider the cases where $r_{\textit{sv}}\in(0,\pi)$.

For the four cases, the conditions that achieve maximal and infimum of viewpoint leakage probabilities can be obtained from \eqref{PL_prob_4_cases} and \eqref{e_4_cases}, which are listed in Table \ref{table:limit_four_regions}. To show the impact of the configured resources on the viewpoint leakage probability, we also provide the monotonicity of $\mathrm{Pr}_{\mathrm{q}}$ w.r.t. $r_{\textit{sv}}$.

\begin{table}[htbp]
	\vspace{-0.75cm}
	\caption{Maximal and infimum of $\mathrm{Pr}_{\mathrm{q}}$ in the four cases when $r_{\textit{sv}}\in(0,\pi)$}\label{table:limit_four_regions}
	\vspace{-0.65cm}
	\begin{center}
		\begin{tabular}{|c|c|c|c|c|c|}
			\hline
			\multirow{2}{*}{Case}&\multicolumn{2}{c|}{ $\mathrm{Pr}_{\mathrm{q}}=1$}&Monotonicity of $\mathrm{Pr}_{\mathrm{q}}$&\multicolumn{2}{c|}{$\mathrm{Pr}_{\mathrm{q}}= \mathrm{Pr}_{\mathrm{q}}^{\inf}$ $^*$}\\
			\cline{2-3}\cline{5-6}
			&\multicolumn{2}{c|}{Conditions w.r.t. $e$ and $r_{\textit{sv}}$} &w.r.t. $r_{\textit{sv}}$ when $\mathrm{Pr}_{\mathrm{q}}<1$ &Value of $\mathrm{Pr}_{\mathrm{q}}^{\inf}$&Condition w.r.t. $r_{\textit{sv}}$$^{\vartriangle}$ \\
			\hline
			\textit{FoV}$\subset$\textit{SFoV}&\multirow{2}{*}{$e\leq\varepsilon$}&\multirow{2}{*}{$|r_{\textit{sv}}-r_{\textit{fov}}|\leq \varepsilon$}&Increasing&$\frac{1 - \cos(\varepsilon)}{1 + \cos(r_{\textit{fov}})}>\mathrm{Pr}^{\min}$&$\pi - r_{\textit{sv}}\leq \delta^{\star}$\\
			\cline{1-1} \cline{4-6}
			\textit{SFoV}$\subset$\textit{FoV}&&&Decreasing&$\frac{1 - \cos(\varepsilon)}{1 - \cos(r_{\textit{fov}})}>\mathrm{Pr}^{\min}$&$r_{\textit{sv}}\leq \delta$\\
			\hline
			\textit{FoV} \!$\cap$\! \textit{SFoV}=$\emptyset$&\multirow{2}{*}{$e\geq\pi-\varepsilon$}&\multirow{2}{*}{$|\pi - r_{\textit{sv}} - r_{\textit{fov}}|\leq\varepsilon$}&Decreasing&$\frac{1 - \cos(\varepsilon)}{1 + \cos(r_{\textit{fov}})}>\mathrm{Pr}^{\min}$&$r_{\textit{sv}}\leq \delta$\\
			\cline{1-1} \cline{4-6}
			\textit{SFoV-C}$\subset$\textit{FoV}&&&Increasing&$\frac{1 - \cos(\varepsilon)}{1 - \cos(r_{\textit{fov}})}>\mathrm{Pr}^{\min}$&$\pi - r_{\textit{sv}}\leq \delta$\\
			\hline
		\end{tabular}
	\end{center}
	\footnotesize{ \ \ \ \ \ \ $^*$ $\mathrm{Pr}_{\mathrm{q}}^{\inf}$ is the infimum of $\mathrm{Pr}_{\mathrm{q}}$. \ \
	$^{\vartriangle}$ $e$ can be an arbitrary value.
	\ \ $^{\star}\delta$ is an arbitrarily small positive number.}
	\vspace{-0.75cm}
\end{table}

One condition for $\mathrm{Pr}_{\mathrm{q}}$ achieving one, $e\leq\varepsilon$ or $e\geq\pi -\varepsilon$, \textit{ indicates that either the predicted viewpoint or its symmetry point is in the $\varepsilon$-viewpoint sensitive neighborhood.} The other condition, $r_z\leq\varepsilon$, where $r_z= |r_{\textit{sv}}-r_{\textit{fov}}|$ or $|\pi - r_{\textit{sv}} - r_{\textit{fov}}|$, \textit{indicates that the configured resources make the difference between the radius of SFoV or SFoV-C and the radius of FoV no larger than $\varepsilon$.}
Specifically, for \textit{FoV}$\subset$\textit{SFoV} and \textit{SFoV}$\subset$\textit{FoV}, $e = \mathtt{d}(O_p, O_v)\leq \varepsilon$. This indicates that the predicted viewpoint lies in the viewpoint sensitive neighborhood. For \textit{FoV} \!$\cap$\! \textit{SFoV}=$\emptyset$ and \textit{SFoV-C}$\subset$\textit{FoV}, the arc length $\mathtt{d}(O_v, \widetilde{O}_p)=\mathtt{d}(O_p, \widetilde{O}_p) - \mathtt{d}(O_p, O_v) = \pi - e \leq \varepsilon$, where $\widetilde{O}_p$ is the symmetry point of $O_p$ w.r.t. the center of the unit sphere.
This indicates that the symmetry point of $O_p$ is in the viewpoint-sensitive neighborhood.
\textit{However, low or high prediction error (i.e., $e\leq \varepsilon$ or $e\geq \pi - \varepsilon$) itself does not lead to $\mathrm{Pr}_{\mathrm{q}}=1$. Only if the conditions of prediction error and configured resources hold simultaneously, $\mathrm{Pr}_{\mathrm{q}}=1$.  }

\textit{When $|r_{\textit{sv}}-r_{\textit{fov}}|> \varepsilon$ or $|\pi - r_{\textit{sv}} - r_{\textit{fov}}|>\varepsilon$, $\mathrm{Pr}_{\mathrm{q}}<1$ and the viewpoint leakage probability decreases as $r_{\textit{sv}}\rightarrow0$ or $r_{\textit{sv}}\rightarrow\pi$}. The
viewpoint leakage probability achieves the global minimum when $r_{\textit{sv}}=0$ or $\pi$, which however is not defined in the four cases. Nevertheless, the infimum still exists, and the infima of viewpoint leakage probabilities in the four cases are larger than the global minimum $\mathrm{Pr}^{\min}$. This is because the radius of possible viewpoint zone $r_z$ is not continuous when $r_{\textit{sv}}=0$ or $\pi$. For example, when \textit{FoV}$\subset$\textit{SFoV}, the limit of $r_z$ from the left is $\lim\limits_{r_{\textit{sv}}\rightarrow\pi^{-}}r_z = \pi - r_{\textit{fov}}$. However, when $r_{\textit{sv}}=\pi$ that is not the case of \textit{FoV}$\subset$\textit{SFoV}, the possibly inferred viewpoint can be arbitrary on the sphere and the radius of possible viewpoint zone becomes $r_z=\pi$.

For the remaining case, since $e$ can be determined from the values of $\mathrm{QoE}$ and $r_{\textit{sv}}$, the viewpoint leakage probability achieves the maximum when $e\in[0, \arcsin(\frac{\varepsilon}{\pi})]\cup[\pi -  \arcsin(\frac{\varepsilon}{\pi}),\pi]$, as shown in \eqref{g_when_PL_prob_circle=1}.
As analyzed in subsection \ref{subsection:VLP_given_e}, $\mathrm{Pr}_{\mathrm{e}}^{\min} = \frac{\varepsilon}{\pi}$, which is achieved when $e=0.5\pi$.

When $\varepsilon\in[0,r_{\textit{fov}}]$ and $r_{\textit{fov}}\in[0,0.5\pi]$, after some regular derivations, we can obtain that $\frac{1 - \cos(\varepsilon)}{2}<\frac{\varepsilon}{\pi}$. This indicates that the minimal viewpoint leakage probability when uploading the QoE metric is less than the minimal viewpoint leakage probability when uploading the prediction error, i.e., $\mathrm{Pr}_{\mathrm{q}}^{\min}<\mathrm{Pr}_{\mathrm{e}}^{\min}$. This is because the real viewpoint is inferred indirectly from the QoE metric via prediction error, which agrees with the intuition.

\vspace{-0.4cm}
\subsection{Relations of Viewpoint Leakage with QoE and Configured Resources}
\begin{figure}[htbp]
	\vspace{-0.7cm}
	\centering
	\subfloat[\textit{FoV}$\subset$\textit{SFoV}, trade resources for privacy.]{\label{Fig:PL_prob_fov_in_cc_numerical}
		\begin{minipage}[c]{0.5\linewidth}
			\centering
			\includegraphics[width=1\textwidth]{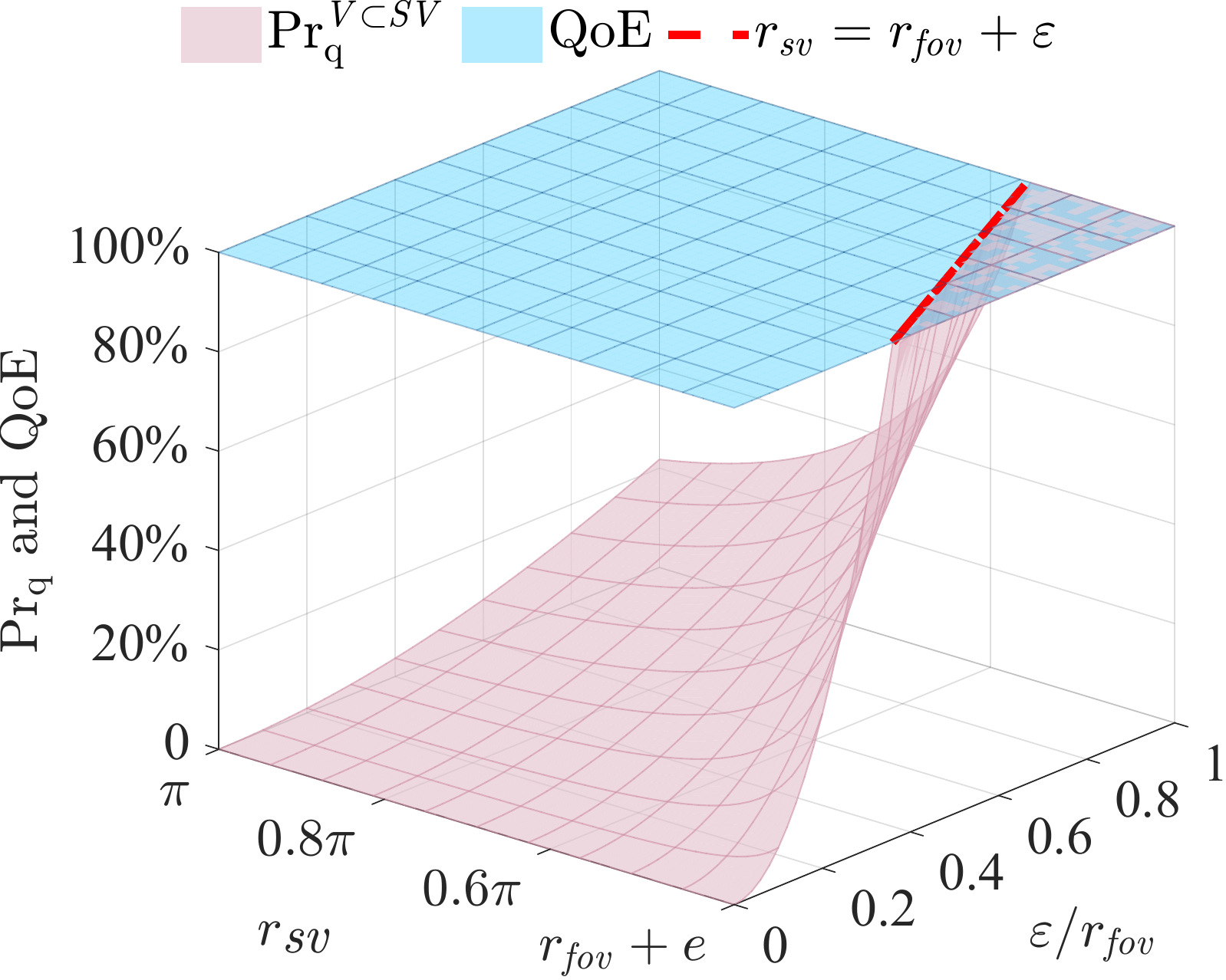}
		\end{minipage}
	}
	\subfloat[\textit{SFoV}$\subset$\textit{FoV},  trade resources and privcay for QoE.]{\label{Fig:PL_prob_cc_in_fov_numerical}
		\begin{minipage}[c]{0.5\linewidth}
			\centering
			\includegraphics[width=1\textwidth]{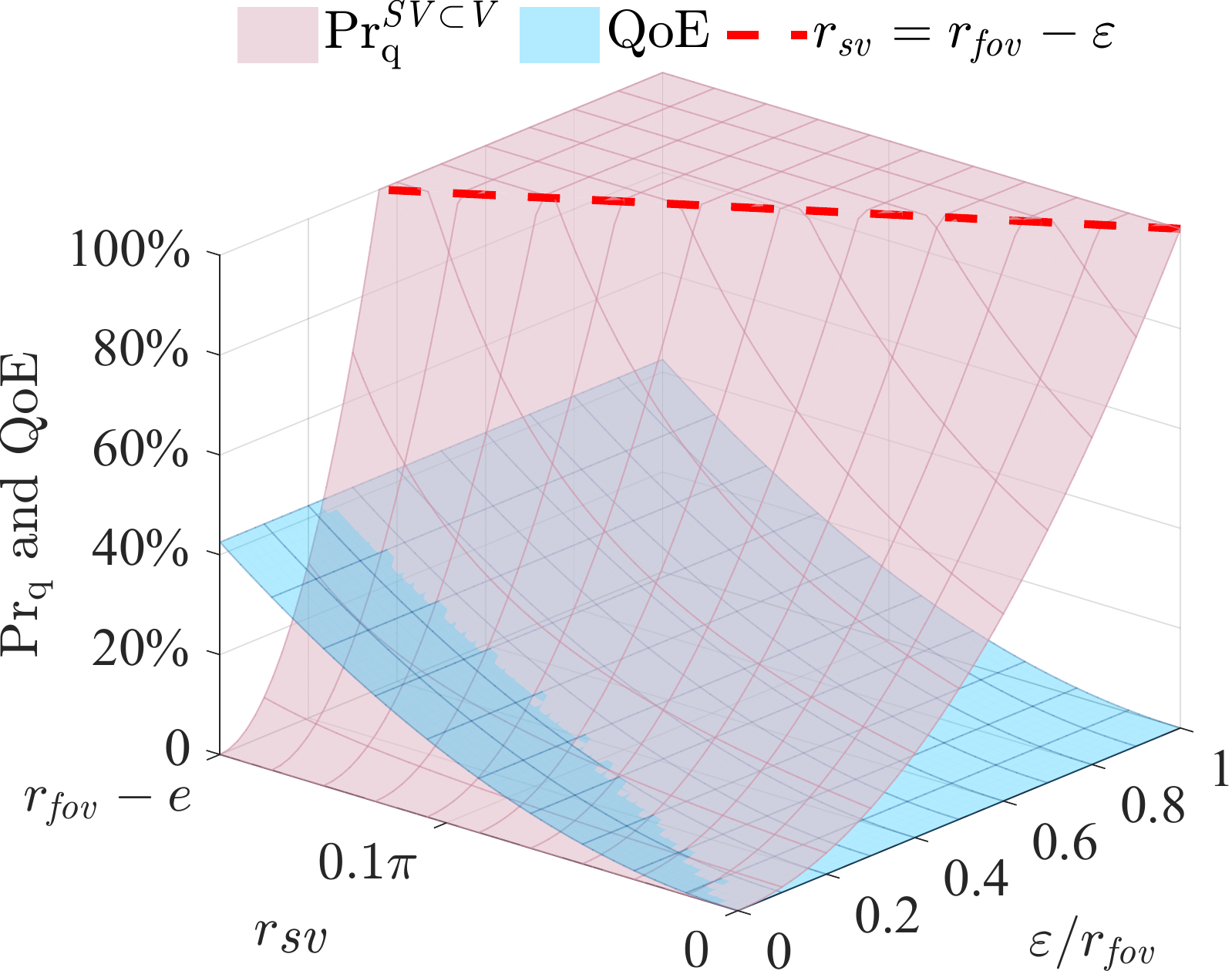}
		\end{minipage}
	}\\
	\subfloat[\textit{FoV} \!$\cap$\! \textit{SFoV}=$\emptyset$, exception because of $\mathrm{QoE}=0$ .]{\label{Fig:PL_prob_fov_cc_no_intersection_numerical}
		\begin{minipage}[c]{0.5\linewidth}
			\centering
			\includegraphics[width=1\textwidth]{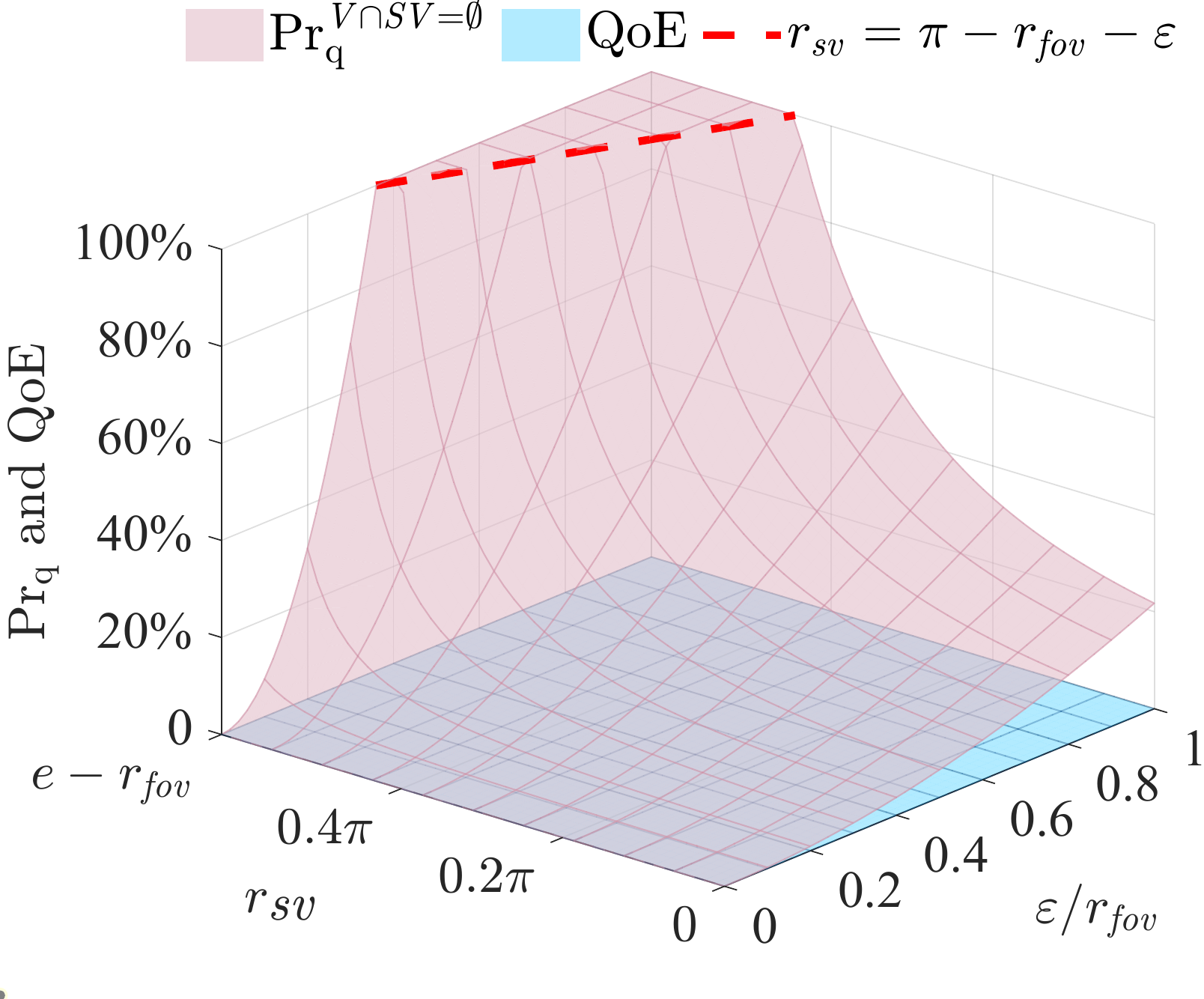}
		\end{minipage}
	}
	\subfloat[\textit{SFoV-C}$\subset$\textit{FoV}, trade resources for QoE and privacy.]{\label{Fig:PL_prob_inverse_cc_in_fov_numerical}
		\begin{minipage}[c]{0.5\linewidth}
			\centering
			\includegraphics[width=1\textwidth]{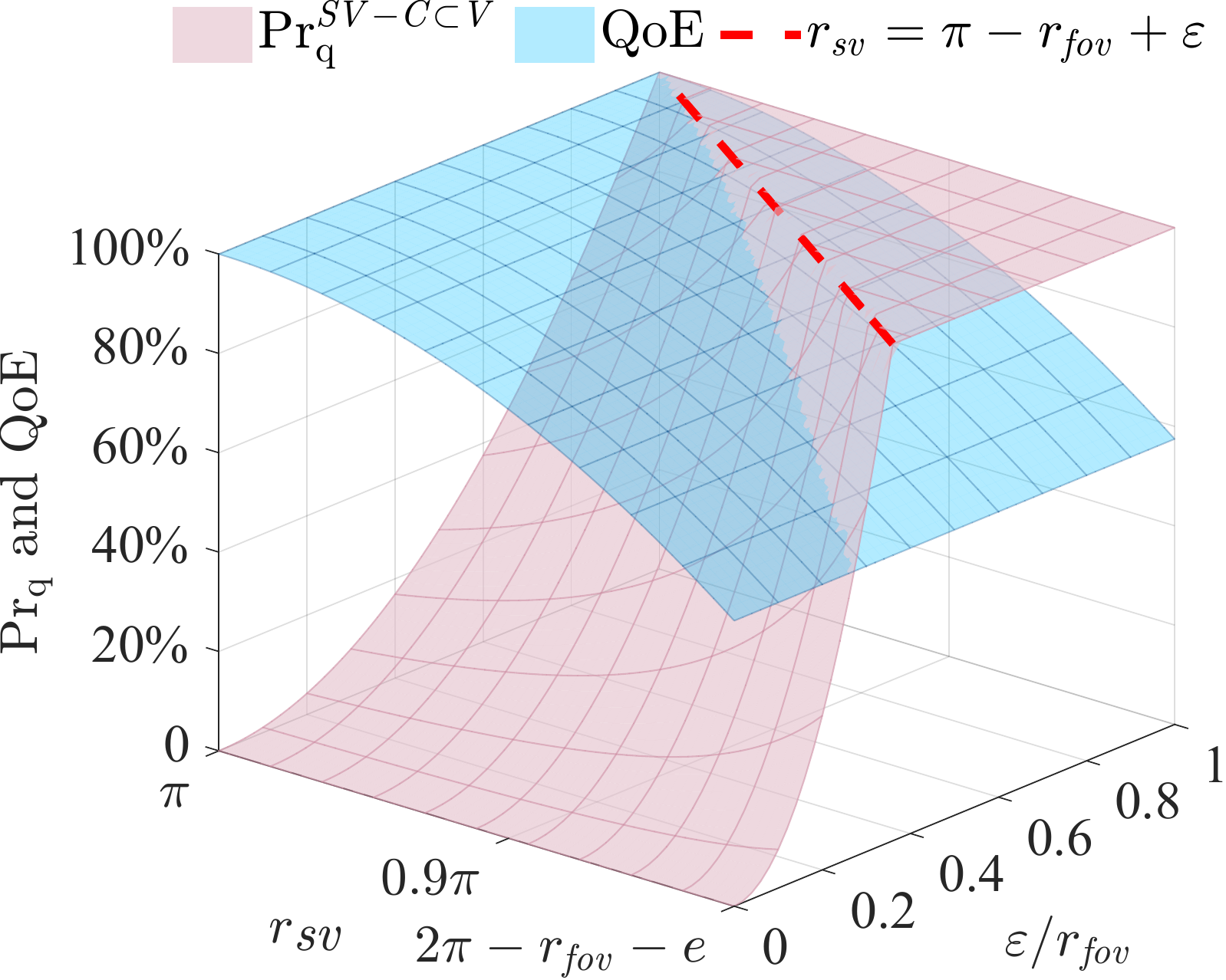}
		\end{minipage}
	}
	\vspace{-0.35cm}
	\caption{$\varepsilon$-viewpoint leakage probability and $\mathrm{QoE}$ in the four cases when $r_{\textit{sv}}\in(0,\pi)$. In Figs. \ref{Fig:PL_prob_fov_in_cc_numerical} and \ref{Fig:PL_prob_cc_in_fov_numerical}, $e=0.1\pi$. In Figs. \ref{Fig:PL_prob_fov_cc_no_intersection_numerical} and \ref{Fig:PL_prob_inverse_cc_in_fov_numerical}, $e=0.9\pi$.}\label{Fig:PL_prob_four_regions_numerical}	
	\vspace{-1.2cm}
\end{figure}

For the four cases, when $|r_{\textit{sv}}-r_{\textit{fov}}|\leq \varepsilon$ or $|\pi - r_{\textit{sv}} - r_{\textit{fov}}|\leq\varepsilon$, $\mathrm{Pr}_{\mathrm{q}}=1$, otherwise $\mathrm{Pr}_\mathrm{q}$ is a monotonic function of $r_{\textit{sv}}$.
When $\mathrm{Pr}_{\mathrm{q}}$ decreases as $r_{\textit{sv}}$ increases, more resources are required  to protect the viewpoint privacy, which also improves the QoE. When $\mathrm{Pr}_{\mathrm{q}}$ decreases as $r_{\textit{sv}}$ decreases, less resources should be configured to protect the viewpoint privacy, which however sacrifices the QoE unless the QoE metric is already zero.
That is to say, \textit{unless $\mathrm{Pr}_{\mathrm{q}}=1$ or $\mathrm{QoE}=0$, either a privacy-resources tradeoff or a privacy-QoE tradeoff exists}.
To visualize the relation, we provide values of $\mathrm{QoE}$ obtained from \eqref{QoE_funct} and $\mathrm{Pr}_\mathrm{q}$ obtained from \eqref{PL_prob_4_cases} in Fig. \ref{Fig:PL_prob_four_regions_numerical}.

\begin{XingRmk}\label{Remark:tradeoff_exception}
	\vspace{-0.3cm}
	We can observe from Fig. \ref{Fig:PL_prob_four_regions_numerical} that neither of the tradeoffs exist (i.e., $\mathrm{Pr}_{\mathrm{q}}=1$ or $\mathrm{QoE}=0$) when (a) $r_{\textit{sv}}\leq r_{\textit{fov}}+ \varepsilon$ in \textit{FoV \!$\subset$\! SFoV}, (b) $r_{\textit{sv}} \geq r_{\textit{fov}} - \varepsilon$ in \textit{SFoV \!$\subset$\! FoV}, (c) $r_{\textit{sv}} \leq \pi - r_{\textit{fov}} +\varepsilon$ in \textit{SFoV-C$\subset$FoV}, (d) in \textit{FoV \!$\cap$\! SFoV}=$\emptyset$. We refer to (a)-(d) as the ``exceptional cases". The conditions for these cases appearing depend on the values of $r_{\textit{sv}}$ and $e$, as shown in \eqref{QoE_funct}.
\end{XingRmk}
\vspace{-0.5cm}

\vspace{-0.1cm}
\section{Trace-Driven Simulation Results}

In this section, with a state-of-the-art predictor, we show that the privacy-QoE
tradeoff always exists when uploading the prediction error, and the exceptional cases where neither the privacy-resources tradeoff nor the privacy-QoE tradeoff exists rarely occur when uploading the QoE metric.

We consider the viewpoint prediction on a real dataset \cite{NTHU_dataset}, where $r_{\textit{fov}}=50^{\circ}$ and 300 traces of viewpoints from 30 users watching 10 VR videos are used for training and testing predictors.\footnote{According to the analysis in  \cite{TRACK}, the traces of the first 20 users in the dataset have mistakes, thus we only use the traces of the other 30 users. Besides, the conclusion obtained in this section is also true for other datasets.}
The total traces are randomly split into training and test sets with ratio 8:2 \cite{TRACK}. The number of traces in test set is $n_{\rm{trace}} = 30\times10\times0.2=60$. For each trace with the playback duration 60 s, the viewpoint is sampled five times per second \cite{TRACK}.
The total proactive streaming time is set as $T_{\mathrm{ps}}=2$ s, the playback duration of a segment is $T_{\mathrm{seg}}=1$ s \cite{Xing_VR_Shannon}. The initial two segments are streamed in the passive manner.
Then, the number of viewpoint samples in each trace to be predicted is $n_{\rm{sp}} = 60\times5 - 2\times 5=290$. The total number of viewpoint samples to be predicted is $N_{\mathrm{sp}}=n_{\rm{trace}}\times n_{\rm{sp}}=17400$.

We use the \textit{deep-position-only} predictor, which achieves the best accuracy for the dataset according to evaluation in \cite{TRACK}.
The predictor employs a sequence-to-sequence long short term memory (LSTM)-based architecture, which uses the viewpoint positions in an observation window as input to predict the positions in a prediction window \cite{TRACK}. In the initial setting of the predictor, the prediction window starts immediately after the observation window. To reserve time for computing and communication, we tailor the predictor by setting the duration between the end of the observation window and the beginning of the prediction windows as $T_{\mathrm{cc}}=1$ s, and set the durations of observation and prediction windows as $T_{\mathrm{obw}}=1$ s and $T_{\mathrm{pdw}} = T_{\mathrm{seg}}=1$ s, respectively. To protect the viewpoint, we consider training and predicting at the HMD, which is case 4 in Table \ref{table:FoV_leakage_one_manner}. When training the predictor, we consider a classical federated learning algorithm, \texttt{FederatedAveraging} \cite{google_federated_learning_17}. The settings of the federated learning are as follows. In each round, the model parameters of the predictor are updated at the HMDs of all of the $K=30$ users. The number of local epochs for each user is $E_l=50$, the number of communication rounds is $R=10$.
The weighting coefficient of the $k$th user on the model parameter is $c_k =\frac{n_k}{N_{\textit{train}}}$, where $N_{\textit{train}}=300\times0.8=240$ is the total number of traces in the training set, and $n_k$ is the number of video traces of the $k$th user in the training set. Due to the random division of training and test sets, $n_k$ varies from 6 to 10. We refer to the predictor as \textit{tailored federated position-only} predictor. Other details and hyper-parameters of the tailored predictor are the same as the \textit{deep-position-only} predictor \cite{TRACK}.
To gain useful insight, we assume that all the users have identical privacy requirements $(\varepsilon, \mathrm{Pr}_{\mathrm{e}}^u)$ for all the videos.
After making prediction, the prediction errors for all the viewpoint samples $\mathrm{E}\triangleq\{e_1,...,e_{N_{\mathrm{sp}}}\}$ can be obtained.

\vspace{-0.5cm}
\subsection{Relations of Viewpoint Privacy with Prediction Performance and QoE When Uploading
$e$}
\begin{figure}[htbp]
	\vspace{-0.75cm}
	\centering
	\subfloat[Average $e$ over $E^u$ v.s. $\varepsilon$ and $\mathrm{Pr}_{\mathrm{e}}^u$]{\label{Fig:Simulation_e_average_e}
		\begin{minipage}[c]{0.5\linewidth}
			\centering
			\includegraphics[width=1\textwidth]{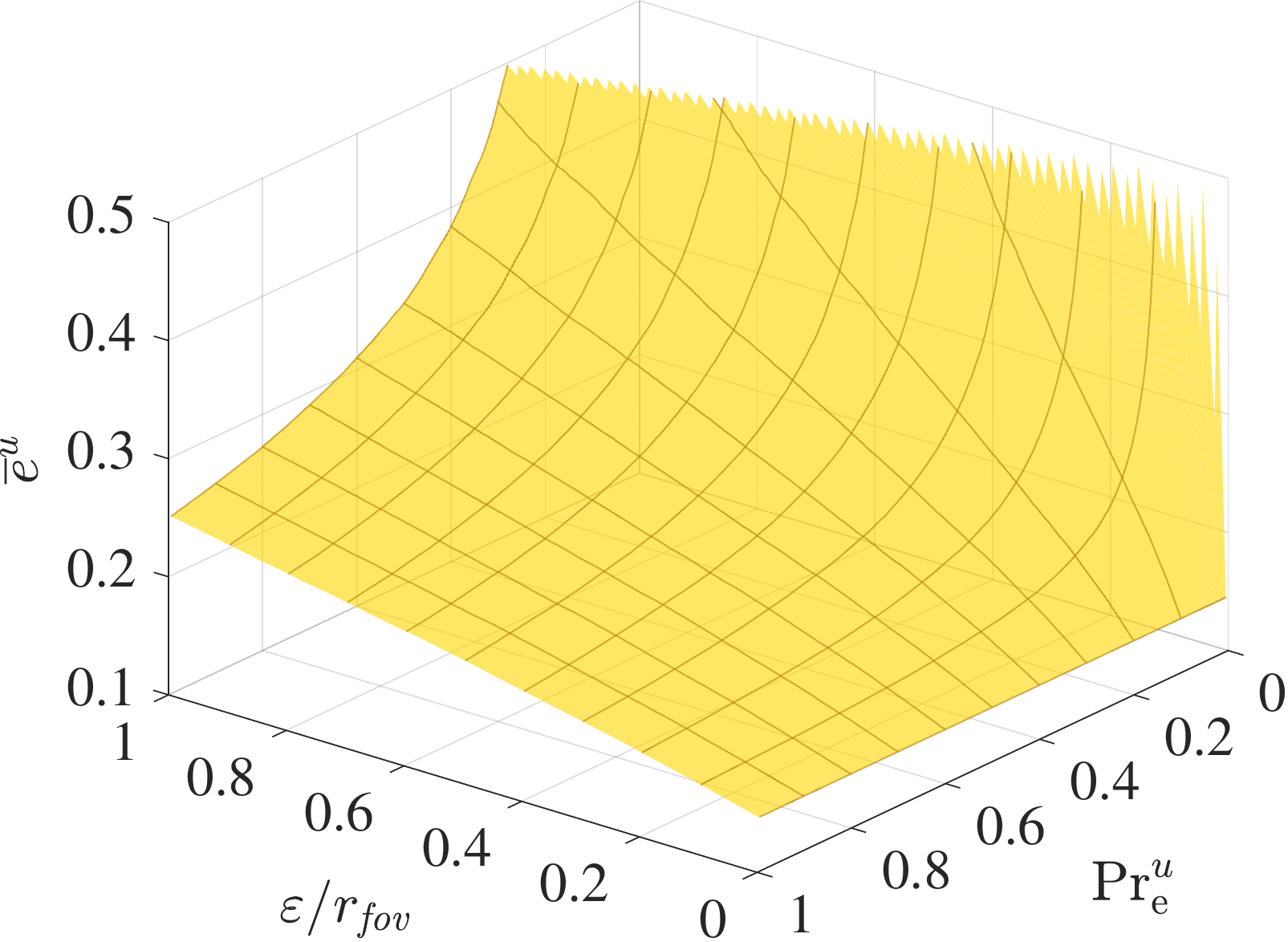}
		\end{minipage}
	}
	\subfloat[Ratios of tradeoff and consistency]{\label{Fig:tradeoff_consist_ratio}
		\begin{minipage}[c]{0.5\linewidth}
			\centering
			\includegraphics[width=1\textwidth]{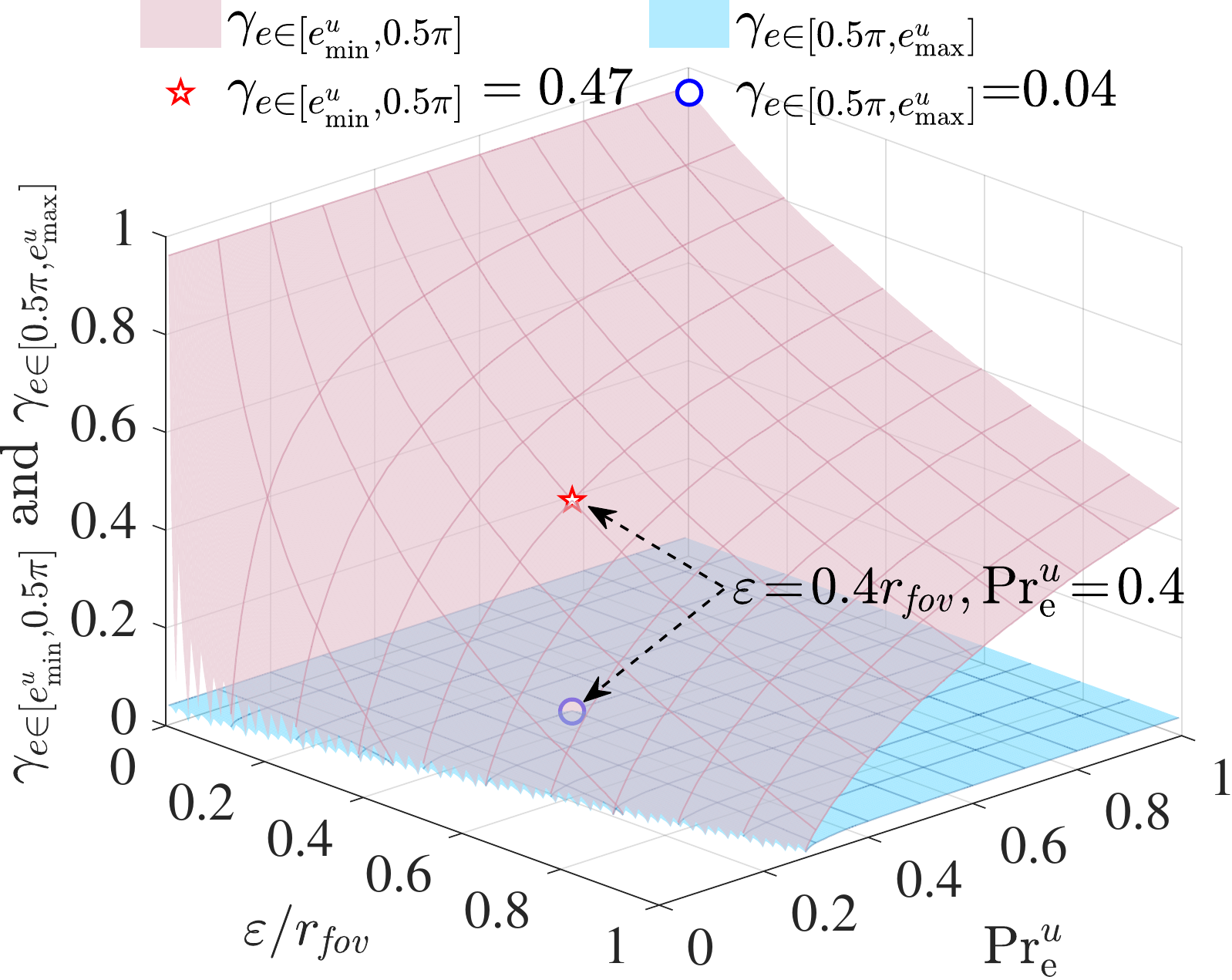}
		\end{minipage}
	}
	\vspace{-0.35cm}
	\caption{Tradeoff and consistency when uploading $e$, $\mathrm{Pr}_{\mathrm{e}}\in[\mathrm{Pr}_{\mathrm{e}}^{\min},1)$ }\label{Fig:Average_viewpoint_leakage_probabilities_upload_e}	
	\vspace{-1cm}
\end{figure}

To evaluate the relation between  the viewpoint privacy requirement and the prediction performance, we first obtain the subset of $E$ that can satisfy the privacy requirement.
According to \eqref{user_req_g},  the subset is $\mathrm{E}^u\triangleq\big\{e_i|e_i\in[e_{\min}^u, e_{\max}^u]\ \textrm{and}\ e_i\in\mathrm{E}\big\}$ given  $\varepsilon\in[0,r_{\textit{fov}}]$ and $\mathrm{Pr}_{\mathrm{e}}^u\in[\mathrm{Pr}_{\mathrm{e}}^{\min},1)$.
Then, we provide the average prediction error over the subset, denoted as $\overline{e}^u$, in Fig. \ref{Fig:Simulation_e_average_e}. We can observe that as the viewpoint privacy requirement becomes more stringent (reflected by increasing $\varepsilon$ or decreasing $\mathrm{Pr}_{\mathrm{e}}^u$), the average prediction error always increases. This implies that the prediction performance of the state-of-the-art predictor should be degraded for a user with stringent privacy requirement. 

Recalling that $\mathrm{QoE}$ is degraded with the increase of prediction error, this indicates that the tradeoff between privacy requirement and QoE improvement always exists.

To explain why the consistency in Fig. \ref{Fig:PL_prob_given_g_2D_numerical} does not appear, we provide the ratios that tradeoff and consistency occur in Fig. \ref{Fig:tradeoff_consist_ratio}, defined as $\gamma_{\textit{tradeoff}}\triangleq\frac{\sum_{i=1}^{N_{\mathrm{sp}}}\mathbbm{1}\left(e_i\in[e_{\min}^u,0.5\pi]\right)}{N_{\mathrm{sp}}}$ and $\gamma_{\textit{consist}}\triangleq \frac{\sum_{i=1}^{N_{\mathrm{sp}}}\mathbbm{1}\left(e_i\in[0.5\pi,e_{\max}^u]\right)}{N_{\mathrm{sp}}}$ according to Remark \ref{Remark:uploading_e_tradeoff}, and $\mathbbm{1}(\cdot)$ is the indicator function.
We can observe that $\gamma_{\textit{tradeoff}}\gg \gamma_{\textit{consist}}$. For example, when $\varepsilon=0.4r_{\textit{fov}}$ and $\mathrm{Pr}_{\mathrm{e}}^u = 0.4$, $\gamma_{\textit{tradeoff}}=0.47$ and $\gamma_{\textit{consist}}=0.04$. This is because a state-of-the-art predictor is employed, and the achieved prediction error is smaller than $0.5\pi$ in high probability.

\vspace{-0.5cm}
\subsection{Relations of Viewpoint Privacy with QoE and Configured Resources When Uploading $\mathrm{\mathrm{QoE}}$}
For the viewpoint leakage probability when uploading the QoE metric, neither of the tradeoffs exist in the exceptional cases mentioned in Remark \ref{Remark:tradeoff_exception}, which depend on the prediction error and configured resources. To evaluate the relations of viewpoint leakage probability with QoE and configured resources, we first analyze the relation of the average viewpoint leakage probability over the prediction errors with the configured resources, and obtain the average QoE over the prediction errors (denoted as $\overline{\mathrm{QoE}}$). Based on which we show that the exceptional cases are significantly reduced.

Given $r_{\textit{sv}}$, the average viewpoint leakage probability over $E$ can be obtained as
\begin{align*}
	\overline{\mathrm{Pr}}_{\mathrm{q}}=
	&\underbrace{\mathrm{Pr}_{\mathrm{q}}^{\textit{V}\subset\textit{SV}}\cdot \gamma^{\textit{V}\!\subset\!\textit{SV}}}_{\triangleq \overline{\mathrm{Pr}}_{\mathrm{q}}^{\textit{V}\subset\textit{SV}} } + \underbrace{\mathrm{Pr}_{\mathrm{q}}^{\textit{SV}\subset\textit{V}}\cdot \gamma^{\textit{SV}\subset\textit{V}}}_{\triangleq\overline{\mathrm{Pr}}_{\mathrm{q}}^{\textit{SV}\subset\textit{V}}} + \underbrace{\mathrm{Pr}_{\mathrm{q}}^{\textit{V} \cap \textit{SV}\!=\!\emptyset}  \cdot \gamma^{\textit{V} \cap \textit{SV}\!=\!\emptyset}}_{\triangleq\overline{\mathrm{Pr}}_{\mathrm{q}}^{\textit{V} \cap \textit{SV}\!=\!\emptyset}} + \underbrace{\mathrm{Pr}_{\mathrm{q}}^{\textit{SV-C}\!\subset\!\textit{V}} \cdot \gamma^{\textit{SV-C}\!\subset\!\textit{V}}}_{\triangleq\overline{\mathrm{Pr}}_{\mathrm{q}}^{\textit{SV-C}\subset\textit{V}}} + \nonumber\\
	&\ \ \ \ \ \ \ \ \ \ \ \ \   \underbrace{\frac{1}{N_{\mathrm{sp}}}\sum_{i=1}^{N_{\mathrm{sp}}} \mathrm{Pr}_{\mathrm{q}}^{\textit{RM}}(e_i)\cdot \mathbbm{1}\left( r_{\textit{sv}}\in\big[r_{\textit{sv},\min}^{rm}(e_i), r_{\textit{sv},\max}^{rm}(e_i)\big] \right)}_{\triangleq \overline{\mathrm{Pr}}_{\mathrm{q}}^{\textit{RM}}},
\end{align*}
where $\gamma^{x}$ is the ratio of case $x$ on $E$. For example, from \eqref{QoE_funct}, $\gamma^{\textit{V}\!\subset\!\textit{SV}}=\frac{1}{N_{\mathrm{sp}}}\sum_{i=1}^{N_{\mathrm{sp}}} \mathbbm{1}(r_{\textit{sv}}\geq r_{\textit{fov}} + e_{i})$. Except the ratios of the four cases, the ratio of the remaining case can be obtained as $\gamma^{\textit{RM}}=\frac{1}{N_{\mathrm{sp}}}\sum_{i=1}^{N_{\mathrm{sp}}} \mathbbm{1}( r_{\textit{sv}}\in\big[r_{\textit{sv},\min}^{rm}(e_i), r_{\textit{sv},\max}^{rm}(e_i)\big] )$.

\begin{figure}[htbp]
	\vspace{-0.55cm}
	\centering
	\subfloat[$\overline{\mathrm{Pr}}_{\mathrm{q}}$ v.s. $r_{\textit{sv}}$ and $\varepsilon$, regions of $r_{\textit{sv}}$: $\mathcal{I}_1$: $(0,r_{\textit{fov}} - \varepsilon)$,
	$\mathcal{D}_2$: $(r_{\textit{fov}} - \varepsilon,r_{\textit{fov}} - \arcsin(\frac{\varepsilon}{\pi}))$, $\mathcal{C}$: $(r_{\textit{fov}} - \arcsin(\frac{\varepsilon}{\pi}), r_{\textit{fov}} + \arcsin(\frac{\varepsilon}{\pi}))$,
	$\mathcal{I}_2$: $(r_{\textit{fov}} + \arcsin(\frac{\varepsilon}{\pi}),r_{\textit{fov}} + \varepsilon), \mathcal{D}_1$: $(r_{\textit{fov}} + \varepsilon,\pi)$.]{\label{Fig:average_PL_vs_C_max_epsilon_3D}
		\begin{minipage}[c]{0.47\linewidth}
			\centering
			\includegraphics[width=1\textwidth]{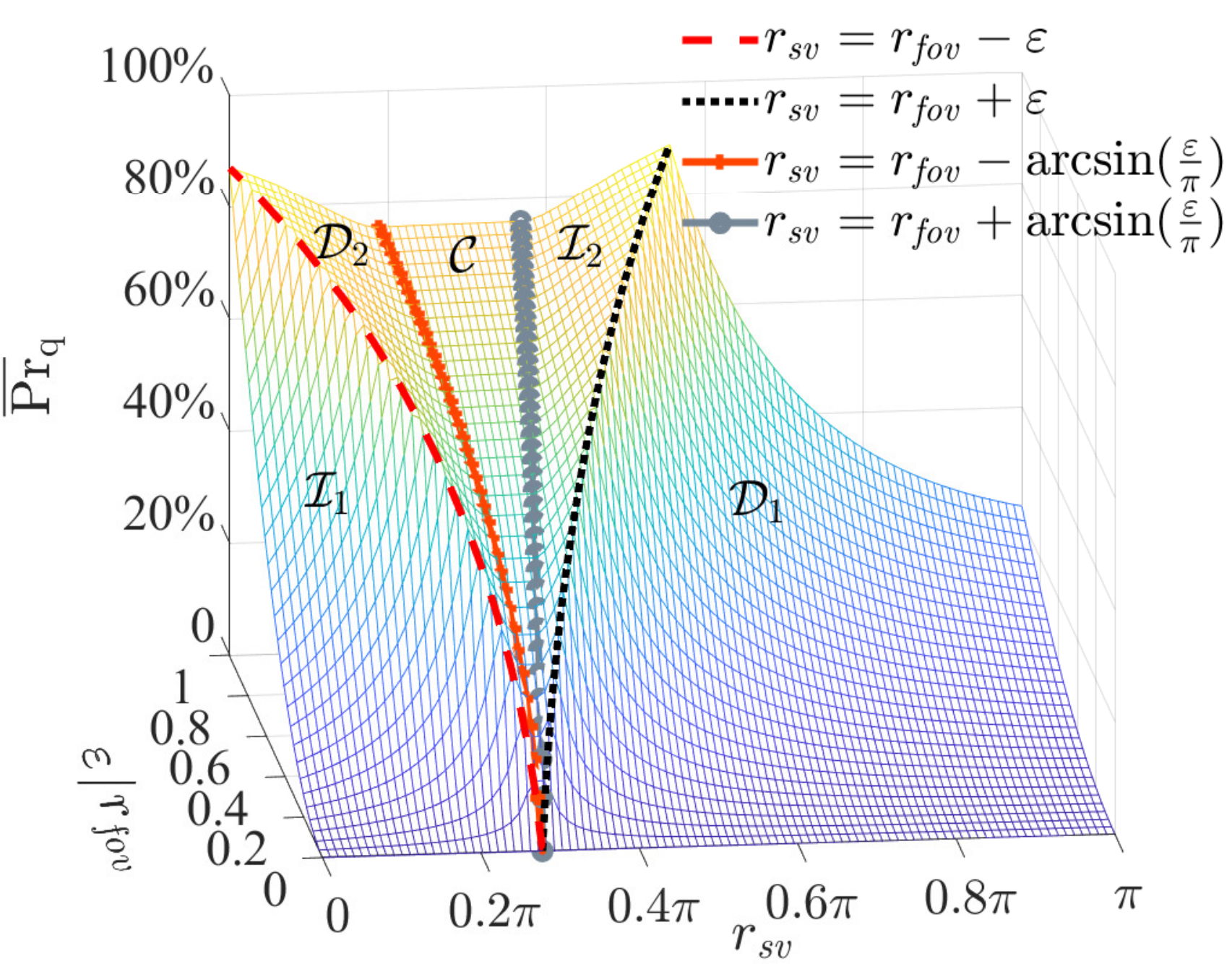}
		\end{minipage}
	}\hspace{2mm}
	\subfloat[$\overline{\mathrm{Pr}}_{\mathrm{q}}$ and $\overline{\mathrm{QoE}}$ v.s. $r_{\textit{sv}}$,  $\mathcal{I}_1$ and $\mathcal{I}_2$: QoE can be improved with more resources at the cost of sacrificing privacy, $\mathcal{D}_1$ and $\mathcal{D}_2$: both QoE and privacy protection can be improved with more resources.]{\label{Fig:average_PL_QoE_vs_r_sv_epsilon04}
	\begin{minipage}[c]{0.47\linewidth}
		\centering
		\includegraphics[width=1\textwidth]{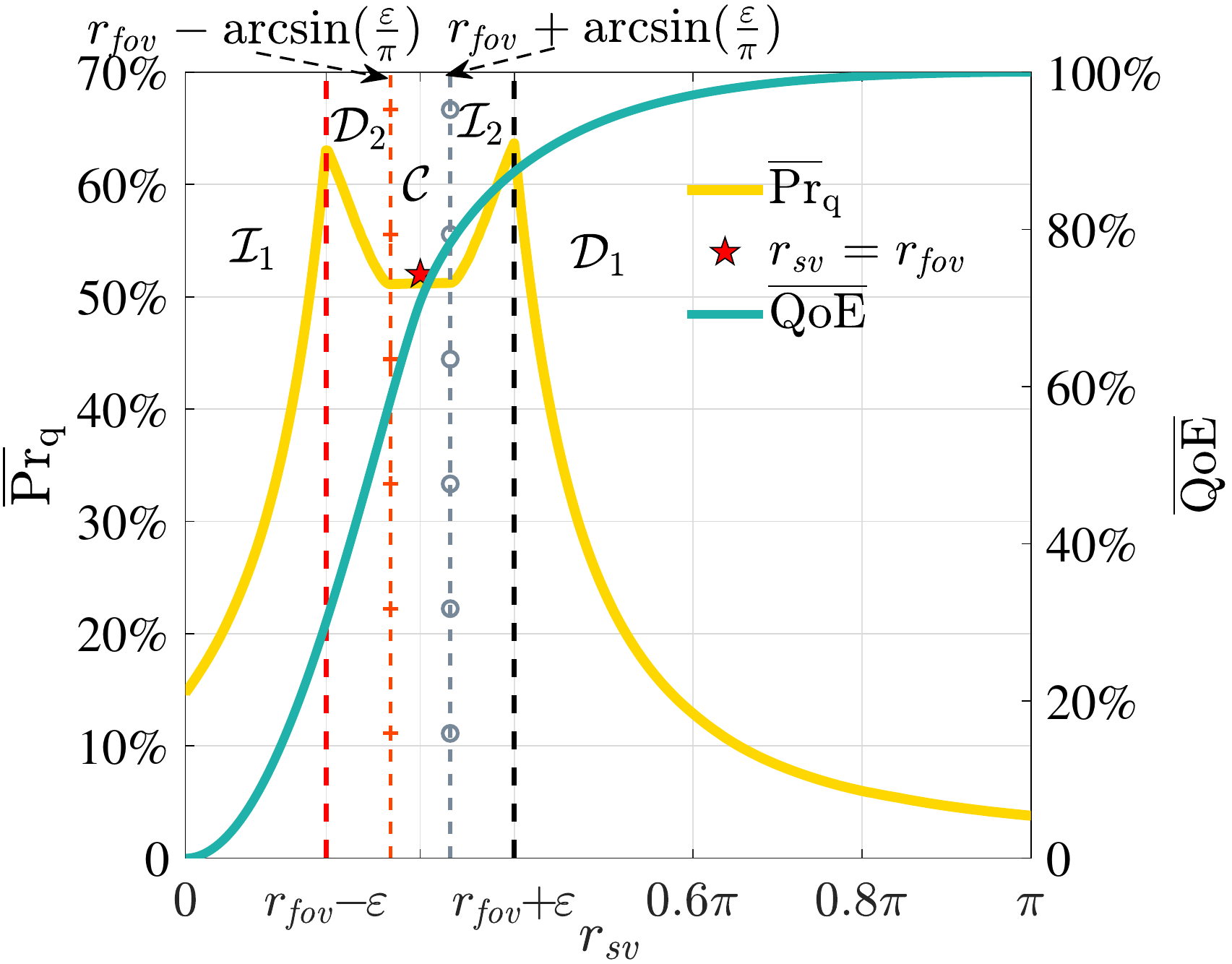}
	\end{minipage}
}\\
	\subfloat[Average viewpoint leakage probabilities v.s. $r_{\textit{sv}}$.
	]{\label{Fig:average_PLs_vs_r_sv_epsilon04}
		\begin{minipage}[c]{0.5\linewidth}
			\centering
			\includegraphics[width=1\textwidth]{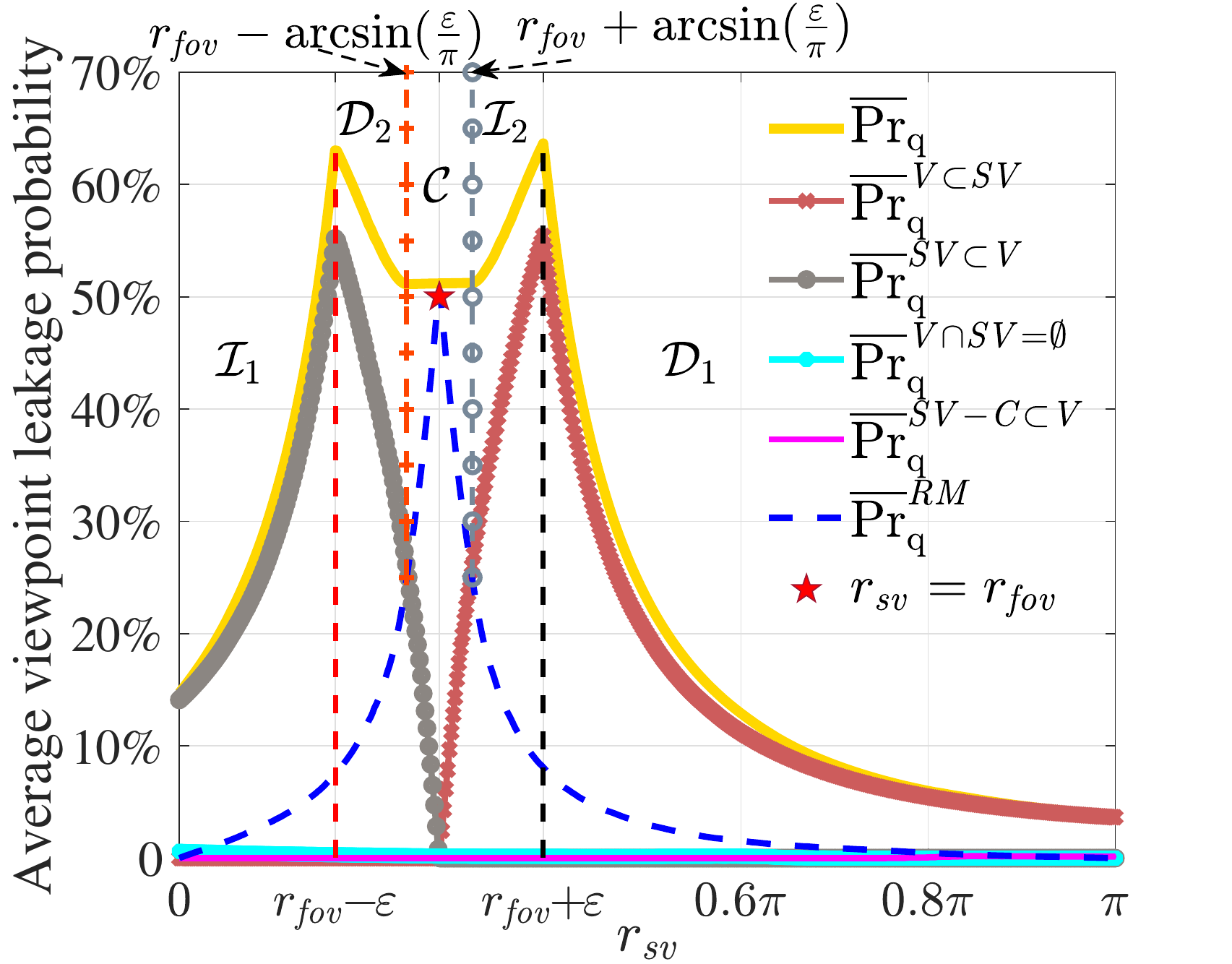}
		\end{minipage}
	}
	\subfloat[Ratio of each case v.s. $r_{\textit{sv}}$.]{\label{Fig:allcases_frequency_vs_r_sv_epsilon04}
		\begin{minipage}[c]{0.5\linewidth}
			\centering
			\includegraphics[width=1\textwidth]{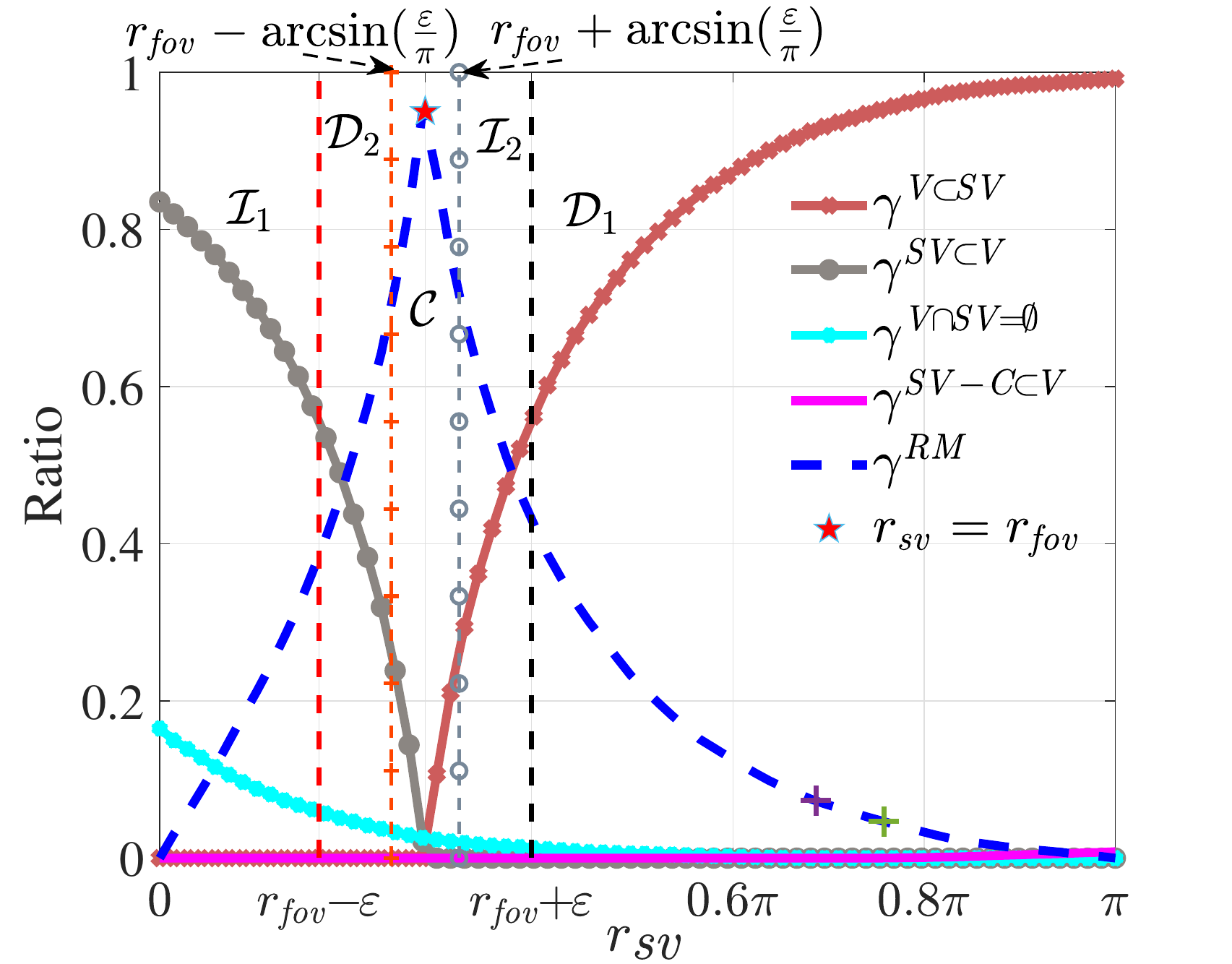}
		\end{minipage}
	}
	\vspace{-0.35cm}
	\caption{Average viewpoint leakage probabilities, $\mathrm{QoE}$, and ratio of each cases when uploading QoE metric, $\varepsilon=0.4r_{\textit{fov}}$ in Fig. (b), (c), and (d).}\label{Fig:Average_viewpoint_leakage_probabilities_upload_QoE}	
	\vspace{-1cm}
\end{figure}

We provide the value of $\overline{\mathrm{Pr}}_{\mathrm{q}}$ in Fig. \ref{Fig:average_PL_vs_C_max_epsilon_3D}. We can observe that given arbitrary $\varepsilon$, the range of $r_{\textit{sv}}$ can be divided in two regions ($\mathcal{I}_1$ and $\mathcal{I}_2$) where $\overline{\mathrm{Pr}}_{\mathrm{q}}$ increases with $r_{\textit{sv}}$, two regions ($\mathcal{D}_1$ and $\mathcal{D}_2$)  where $\overline{\mathrm{Pr}}_{\mathrm{q}}$ decreases with $r_{\textit{sv}}$, and a constant region ($r_{\textit{sv}}\in\mathcal{C}:(r_{\textit{fov}} - \arcsin(\frac{\varepsilon}{\pi}), r_{\textit{fov}} + \arcsin(\frac{\varepsilon}{\pi}))$) where the configured resources make the difference between the radius of SFoV and FoV no larger than $\arcsin(\frac{\varepsilon}{\pi})$.  Since $\overline{\mathrm{Pr}}_{\mathrm{q}}$ is a monotonic function of $r_{\textit{sv}}$ unless $r_{\textit{sv}}\in\mathcal{C}$, we can infer that the exceptional cases only lie in $r_{\textit{sv}}\in\mathcal{C}$. This is verified in Fig. \ref{Fig:average_PL_QoE_vs_r_sv_epsilon04}. Since the relation of $\overline{\mathrm{Pr}}_{\mathrm{q}}$ with $r_{\textit{sv}}$ is similar for other values of $\varepsilon$ as shown in Fig. \ref{Fig:average_PL_vs_C_max_epsilon_3D}, \textit{the exceptional cases happen only when the difference between the radius of SFoV and FoV is no larger than $\arcsin(\frac{\varepsilon}{\pi})$}.

To explain why the exceptional cases are reduced from cases (a)-(d) in Remark \ref{Remark:tradeoff_exception} to $r_{\textit{sv}}\in\mathcal{C}$, we provide the values of average viewpoint leakage probabilities in Fig. \ref{Fig:average_PLs_vs_r_sv_epsilon04} and the ratio of each case in Fig. \ref{Fig:allcases_frequency_vs_r_sv_epsilon04}.

As shown in Fig. \ref{Fig:average_PLs_vs_r_sv_epsilon04}, when the exceptional cases (a) and (b) in Remark \ref{Remark:tradeoff_exception} appear,
$\overline{\mathrm{Pr}}_{\mathrm{q}}^{\textit{V} \subset\textit{SV}}$,  $\overline{\mathrm{Pr}}_{\mathrm{q}}^{\textit{SV} \subset\textit{V}}$, and $\overline{\mathrm{Pr}}_{\mathrm{q}}^{\textit{RM}}$ monotonically increases, decreases, first increases then decreases with $r_{\textit{sv}}$, respectively, although the value of $r_{\textit{sv}}$ does not affect $\mathrm{Pr}_{\mathrm{q}}$.
From Fig. \ref{Fig:allcases_frequency_vs_r_sv_epsilon04} we can observe that this is because the monotonicity of the ratios $\gamma^{\textit{V} \subset\textit{SV}}$, $\gamma^{\textit{SV} \subset\textit{V}}$, and $\gamma^{\textit{RM}}$ versus $r_{\textit{sv}}$, respectively. Only when $r_{\textit{sv}}\in\mathcal{C}$, the increase and decrease of $\overline{\mathrm{Pr}}_{\mathrm{q}}^{\textit{RM}}$ are respectively counteracted by the decrease of  $\overline{\mathrm{Pr}}_{\mathrm{q}}^{\textit{SV}\subset\textit{V}}$ and the increase of  $\overline{\mathrm{Pr}}_{\mathrm{q}}^{\textit{V}\subset\textit{SV}}$.
That is to say, the exceptional cases (a) and (b) in Remark \ref{Remark:tradeoff_exception} is shrunk to $r_{\textit{sv}}\in\mathcal{C}$ because of the relation of $\gamma^{\textit{V} \subset\textit{SV}}$, $\gamma^{\textit{SV} \subset\textit{V}}$, and $\gamma^{\textit{RM}}$ with $r_{\textit{sv}}$ for the given predictor.

As shown in Fig. \ref{Fig:allcases_frequency_vs_r_sv_epsilon04}, the values of $\gamma^{\textit{SV-C}\subset\textit{V}}$ and $\gamma^{\textit{V}\cap\textit{SV}=\emptyset}$ are small. This is because the two cases happens when the prediction error is large according to \eqref{e_4_cases}, while most prediction errors are small with the state-of-the art predictor, as shown in Fig. \ref{Fig:tradeoff_consist_ratio}. That is to say, the exceptional cases (c) and (d) in Remark \ref{Remark:tradeoff_exception} rarely occur because prediction errors for most samples are small.

We can find from Fig. \ref{Fig:average_PL_vs_C_max_epsilon_3D} and \ref{Fig:average_PL_QoE_vs_r_sv_epsilon04} that to reduce the average viewpoint leakage probability, either the average QoE has to be sacrificed or more resources are required unless $r_{\textit{sv}}\in\mathcal{C}$.

\section{Conclusion}
In this paper, we studied viewpoint leakage issue during proactive VR streaming, and found that existing privacy-preserving approaches cannot avoid the leakage.
We defined and derived the viewpoint leakage probability when prediction error or QoE metric is uploaded for adaptive streaming.
When uploading the error of predicting viewpoint, if the error is no larger than half of its maximum, there is a tradeoff between protecting viewpoint privacy and improving QoE, otherwise improving privacy protection is consistent with improving QoE. The targets of maximizing QoE and satisfying viewpoint privacy requirement are always contradictory. The
leakage probability achieves its minimum when the prediction error is the half of its maximum.
When uploading the QoE metric and from which the range of prediction error can be inferred, either the privacy-QoE or the privacy-resources tradeoff exists unless the leakage probability is one or the QoE metric is zero. To reduce the leakage probability, either QoE has to be sacrificed or more resources should be configured.
The leakage probability achieves one when the following two conditions hold simultaneously. (1) Either the predicted viewpoint or its symmetry point is in the viewpoint-sensitive neighborhood. (2) The difference between the radius of SFoV or SFoV-C and the radius of FoV is no larger than the radius of viewpoint-sensitive neighborhood. 
The minimal viewpoint leakage probability when uploading the prediction
error is larger than the minimal viewpoint leakage probability when uploading the QoE metric. 

Simulation results with the state-of-the-art predictor over a real dataset show that when uploading the prediction error, the tradeoff between viewpoint privacy and QoE improvement always exists. To reduce the viewpoint leakage probability, the prediction performance should be degraded rather than improved. When uploading the QoE metric, the cases where neither the privacy-resources tradeoff nor the privacy-QoE tradeoff exist rarely occur. To reduce the average viewpoint leakage probability, either the average QoE has to be sacrificed or more resources are required.

\bibliographystyle{IEEEtranTCOM}
\bibliography{IEEEabrv_Xing,privacy_leak_VR_ref}

\begin{thebibliography}{10}
\baselineskip 12pt
\providecommand{\url}[1]{#1}
\csname url@samestyle\endcsname
\providecommand{\newblock}{\relax}
\providecommand{\bibinfo}[2]{#2}
\providecommand{\BIBentrySTDinterwordspacing}{\spaceskip=0pt\relax}
\providecommand{\BIBentryALTinterwordstretchfactor}{4}
\providecommand{\BIBentryALTinterwordspacing}{\spaceskip=\fontdimen2\font plus
\BIBentryALTinterwordstretchfactor\fontdimen3\font minus
  \fontdimen4\font\relax}
\providecommand{\BIBforeignlanguage}[2]{{%
\expandafter\ifx\csname l@#1\endcsname\relax
\typeout{** WARNING: IEEEtran.bst: No hyphenation pattern has been}%
\typeout{** loaded for the language `#1'. Using the pattern for}%
\typeout{** the default language instead.}%
\else
\language=\csname l@#1\endcsname
\fi
#2}}
\providecommand{\BIBdecl}{\relax}
\BIBdecl

\bibitem{survey_Hsu}
C.-L. Fan, W.-C. Lo, Y.-T. Pai, and C.-H. Hsu, ``A survey on 360$^\circ$ video
  streaming: Acquisition, transmission, and display,'' \emph{ACM Comput.
  Surv.}, vol.~52, no.~4, Aug. 2019.

\bibitem{VR_IoT}
C.~Chaccour, M.~N. Soorki, W.~Saad, M.~Bennis, and P.~Popovski, ``Can terahertz
  provide high-rate reliable low latency communications for wireless {VR}?''
  \emph{IEEE Internet Things J.}, early access, 2022.

\bibitem{NTHU_dataset}
W.-C. Lo, C.-L. Fan, J.~Lee, C.-Y. Huang, K.-T. Chen, and C.-H. Hsu,
  ``360$^{\circ}$ video viewing dataset in head-mounted virtual reality,''
  \emph{ACM MMSys}, 2017.

\bibitem{optimizing_VR}
F.~Qian, L.~Ji, B.~Han, and V.~Gopalakrishnan, ``Optimizing 360 video delivery
  over cellular networks,'' \emph{ACM SIGCOMM Workshop}, 2015.

\bibitem{VR_TWC}
L.~Teng, G.~Zhai, Y.~Wu, X.~Min, W.~Zhang, Z.~Ding, and C.~Xiao, ``{QoE} driven
  {VR} 360$^{\circ}$ video massive {MIMO} transmission,'' \emph{{IEEE} Trans.
  Wireless Commun.}, vol.~21, no.~1, pp. 18--33, 2022.

\bibitem{VR_mobile_computing}
L.~Zhong, X.~Chen, C.~Xu, Y.~Ma, M.~Wang, Y.~Zhao, and G.-M. Muntean, ``A
  multi-user cost-efficient crowd-assisted {VR} content delivery solution in
  {5G}-and-beyond heterogeneous networks,'' \emph{{IEEE} Trans. Mobile
  Comput.}, early access, 2022.

\bibitem{yingcui_TWC}
C.~Guo, L.~Zhao, Y.~Cui, Z.~Liu, and D.~W.~K. Ng, ``Power-efficient wireless
  streaming of multi-quality tiled 360 {VR} video in {MIMO-OFDMA} systems,''
  \emph{{IEEE} Trans. Wireless Commun.}, vol.~20, no.~8, pp. 5408--5422, 2021.

\bibitem{VR_broadcast_TWC}
F.~Hu, Y.~Deng, and A.~Hamid~Aghvami, ``Cooperative multigroup broadcast
  360$^{\circ}$ video delivery network: A hierarchical federated deep
  reinforcement learning approach,'' \emph{{IEEE} Trans. Wireless Commun.},
  early access, 2021.

\bibitem{VR_transcoding_JSAC}
H.~Xiao, C.~Xu, Z.~Feng, R.~Ding, S.~Yang, L.~Zhong, J.~Liang, and G.-M.
  Muntean, ``A transcoding-enabled 360$^{\circ}$ {VR} video caching and
  delivery framework for edge-enhanced next-generation wireless networks,''
  \emph{{IEEE} J. Select. Areas Commun.}, early access, 2022.

\bibitem{VR_letter}
Z.~Gu, H.~Lu, and C.~Zou, ``Horizontal and vertical collaboration for {VR}
  delivery in {MEC}-enabled small-cell networks,'' \emph{{IEEE} Commun. Lett.},
  vol.~26, no.~3, pp. 627--631, 2022.

\bibitem{changyangshe_VR_globecom}
S.~Li, C.~She, Y.~Li, and B.~Vucetic, ``Constrained deep reinforcement learning
  for low-latency wireless {VR} video streaming,'' \emph{IEEE GLOBECOM}, 2021.

\bibitem{junnizou_TCSVT}
N.~Kan, J.~Zou, C.~Li, W.~Dai, and H.~Xiong, ``Rapt360: Reinforcement
  learning-based rate adaptation for 360-degree video streaming with adaptive
  prediction and tiling,'' \emph{{IEEE} Trans. Circuits Syst. Video Technol.},
  vol.~32, no.~3, pp. 1607--1623, 2022.

\bibitem{Xueshihou_openjournal}
X.~Hou and S.~Dey, ``Motion prediction and pre-rendering at the edge to enable
  ultra-low latency mobile {6DoF} experiences,'' \emph{IEEE Open Journal of the
  Communications Society}, vol.~1, pp. 1674--1690, 2020.

\bibitem{TRACK}
\BIBentryALTinterwordspacing
M.~F. Romero~Rondon, L.~Sassatelli, R.~Aparicio-Pardo, and F.~Precioso,
  ``{TRACK}: A new method from a re-examination of deep architectures for head
  motion prediction in 360-degree videos,'' \emph{IEEE Trans. Pattern Anal.
  Machine Intell., early access}, 2021. [Online]. Available:
  \url{https://gitlab.com/miguelfromeror/head-motion-prediction/tree/master/}
\BIBentrySTDinterwordspacing

\bibitem{VR_learning_based_TWC}
X.~Liu and Y.~Deng, ``Learning-based prediction, rendering and association
  optimization for {MEC}-enabled wireless virtual reality ({VR}) networks,''
  \emph{{IEEE} Trans. Wireless Commun.}, vol.~20, no.~10, pp. 6356--6370, 2021.

\bibitem{VR_RIS_JSAC}
X.~Liu, Y.~Deng, C.~Han, and M.~D. Renzo, ``Learning-based prediction,
  rendering and transmission for interactive virtual reality in {RIS}-assisted
  terahertz networks,'' \emph{{IEEE} J. Select. Areas Commun.}, vol.~40, no.~2,
  pp. 710--724, 2022.

\bibitem{LSTM_update}
C.~{Perfecto}, M.~S. {Elbamby}, J.~{Del Ser}, and M.~{Bennis}, ``Taming the
  latency in multi-user {VR} 360{\textdegree}: {A} {QoE}-aware deep
  learning-aided multicast framework,'' \emph{IEEE Trans. Commun.}, vol.~68,
  no.~4, pp. 2491--2508, 2020.

\bibitem{QoE_feedback_2}
G.~Xiao, M.~Wu, Q.~Shi, Z.~Zhou, and X.~Chen, ``{DeepVR}: Deep reinforcement
  learning for predictive panoramic video streaming,'' \emph{{IEEE} Trans.
  Cogn. Commun. Netw.}, vol.~5, no.~4, pp. 1167--1177, 2019.

\bibitem{JSAC_private_VR}
R.~Zhang, J.~Liu, F.~Liu, T.~Huang, Q.~Tang, S.~Wang, and F.~R. Yu,
  ``Buffer-aware virtual reality video streaming with personalized and private
  viewport prediction,'' \emph{{IEEE} J. Select. Areas Commun.}, vol.~40,
  no.~2, pp. 694--709, 2022.

\bibitem{Xing_VR_Shannon}
X.~Wei, C.~Yang, and S.~Han, ``Prediction, communication, and computing
  duration optimization for {VR} video streaming,'' \emph{{IEEE} Trans.
  Commun.}, vol.~69, no.~3, pp. 1947--1959, 2021.

\bibitem{VR_survey_prediction}
A.~Yaqoob, T.~Bi, and G.-M. Muntean, ``A survey on adaptive $360^\circ$ video
  streaming: Solutions, challenges and opportunities,'' \emph{IEEE
  Communications Surveys Tutorials}, vol.~22, no.~4, pp. 2801--2838, 2020.

\bibitem{privacy_VR_identifiability}
M.~R. Miller, F.~Herrera, H.~Jun, J.~A. Landay, and J.~N. Bailenson, ``Personal
  identifiability of user tracking data during observation of 360-degree {VR}
  video,'' \emph{Scientific Reports}, vol.~10, no.~1, pp. 1--10, 2020.

\bibitem{privacy-preserving_eye_tracking_2021}
B.~David-John, D.~Hosfelt, K.~Butler, and E.~Jain, ``A privacy-preserving
  approach to streaming eye-tracking data,'' \emph{IEEE Trans. Vis. Comput.
  Graphics}, vol.~27, no.~5, pp. 2555--2565, 2021.

\bibitem{privacy_def_eye_track}
J.~Li, A.~R. Chowdhury, K.~Fawaz, and Y.~Kim, ``Kal$\epsilon$ido: Real-time
  privacy control for eye-tracking systems,'' \emph{30th {USENIX} Security
  Symposium}, Aug. 2021.

\bibitem{Privacy_Preserving_Gaze_Estimation}
E.~Bozkir, A.~B. \"{U}nal, M.~Akg\"{u}n, E.~Kasneci, and N.~Pfeifer, ``Privacy
  preserving gaze estimation using synthetic images via a randomized encoding
  based framework,'' \emph{ACM Symposium on Eye Tracking Research and
  Applications}, 2020.

\bibitem{Differential_Privacy}
A.~Liu, L.~Xia, A.~Duchowski, R.~Bailey, K.~Holmqvist, and E.~Jain,
  ``Differential privacy for eye-tracking data,'' \emph{ACM Symposium on Eye
  Tracking Research and Applications}, 2019.

\bibitem{PrivacEye}
J.~Steil, M.~Koelle, W.~Heuten, S.~Boll, and A.~Bulling, ``Privaceye:
  Privacy-preserving head-mounted eye tracking using egocentric scene image and
  eye movement features,'' \emph{ACM Symposium on Eye Tracking Research and
  Applications}, 2019.

\bibitem{Fixation_Prediction}
C.~{Fan}, S.~{Yen}, C.~{Huang}, and C.~{Hsu}, ``Optimizing fixation prediction
  using recurrent neural networks for 360$^{\circ }$ video streaming in
  head-mounted virtual reality,'' \emph{IEEE Trans. Multimedia}, vol.~22,
  no.~3, pp. 744--759, March 2020.

\bibitem{transmission_mode-standard-update}
3GPP, ``Extended reality ({XR}) in {5G},'' 2020, 3GPP TR 26.928 version 16.0.0
  release 16.

\bibitem{HuaWei_Cloud_VR}
\BIBentryALTinterwordspacing
iLab, ``Cloud {VR} network solution whitepaper,'' Huawei Technologies CO.,
  LTD., Tech. Rep., 2018. [Online]. Available:
  \url{https://www.huawei.com/minisite/pdf/ilab/cloud_vr_network_solution_white_paper_en.pdf}
\BIBentrySTDinterwordspacing

\bibitem{junnizou_TSTSP}
J.~Zou, C.~Li, C.~Liu, Q.~Yang, H.~Xiong, and E.~Steinbach, ``Probabilistic
  tile visibility-based server-side rate adaptation for adaptive 360-degree
  video streaming,'' \emph{{IEEE} J. Sel. Topics Signal Process.}, vol.~14,
  no.~1, pp. 161--176, 2020.

\bibitem{Xing_viewpoint_leakage_arxiv}
X.~Wei, C.~Yang, and C.~Sun, ``Viewpoint leakage in proactive {VR} streaming:
  Modeling and tradeoff,'' \emph{arXiv:2203.03107}.

\bibitem{google_federated_learning_17}
B.~McMahan, E.~Moore, D.~Ramage, S.~Hampson, and B.~A. y~Arcas,
  ``Communication-efficient learning of deep networks from decentralized
  data,'' \emph{PMLR AISTATS}, 2017.

\end{thebibliography}

\end{document}